\documentclass[a4paper,fleqn,usenatbib]{mnras}

\usepackage[T1]{fontenc}
\usepackage{ae,aecompl}
\usepackage[scientific-notation=true]{siunitx}
\DeclareSIUnit\parsec{pc}
\DeclareSIUnit\erg{erg}

\usepackage{graphicx}	% Including figure files
\usepackage{amsmath}	% Advanced maths commands
\usepackage{amssymb}	% Extra maths symbols
\usepackage{ulem}
\usepackage{amsbsy}
\usepackage{mathrsfs,bm}
\usepackage{times}

\newcommand{\bcdot}{\ensuremath{%
  \mathchoice%

  {\mskip\thinmuskip\lower0.2ex\hbox{\scalebox{1.5}{$\cdot$}}\mskip\thinmuskip}}%
   {\mskip\thinmuskip\lower0.2ex\hbox{\scalebox{1.5}{$\cdot$}}\mskip\thinmuskip}%        
   {\lower0.3ex\hbox{\scalebox{1.2}{$\cdot$}}}%  
   {\lower0.3ex\hbox{\scalebox{1.2}{$\cdot$}}}%
}

\newcommand{\bnabla}{\ensuremath{\boldsymbol{\nabla}}}
\newcommand{\bdot}{\ensuremath{\boldsymbol{\cdot}}}
\newcommand{\vect}[1]{\boldsymbol{#1}}

\usepackage{hyperref}
\usepackage[dvipsnames]{xcolor}
\usepackage{array}

\title[Magnetised clusters and velocity fields]
      {Connecting turbulent velocities and magnetic fields in galaxy cluster simulations with active galactic nuclei jets}
\author[K. Ehlert et al.]{%
K. Ehlert$^{1}$\thanks{E-mail: kehlert@aip.de},
R. Weinberger$^{2}$,
C. Pfrommer$^{1}$,
V. Springel$^{3}$
\vspace*{0.2cm}\\
$^{1}$Leibniz Institute for Astrophysics, An der Sternwarte 16, D-14482 Potsdam, Germany\\
$^{2}$Center for Astrophysics | Harvard \& Smithsonian, 60 Garden Street, Cambridge, MA 02138, USA\\
$^{3}$Max-Planck-Institut f\"ur Astrophysik, Karl-Schwarzschild-Str. 1, D-85741 Garching, Germany\\
}
\date{Accepted XXX. Received YYY; in original form ZZZ}
\pubyear{2019}

\begin{document}
\label{firstpage}
\pagerange{\pageref{firstpage}--\pageref{lastpage}}
\maketitle

% Abstract of the paper
\begin{abstract}
  The study of velocity fields of the hot gas in galaxy clusters can help to unravel details of microphysics on small-scales and to decipher the nature of feedback by active galactic nuclei (AGN). Likewise, magnetic fields as traced by Faraday rotation measurements (RMs) inform about their impact on gas dynamics as well as on cosmic ray production and transport. We investigate the inherent relationship between large-scale gas kinematics and magnetic fields through non-radiative magnetohydrodynamical
  simulations of the creation, evolution and disruption of AGN jet-inflated lobes in an isolated Perseus-like galaxy cluster, with and without pre-existing turbulence.  In particular, we connect cluster velocity measurements with mock RM maps to highlight their underlying physical connection, which opens up the possibility of comparing turbulence levels in two different observables. For single jet outbursts, we find only a  local impact on the velocity field, i.e.~the associated increase in velocity dispersion is not volume-filling. Furthermore, in a setup with pre-existing turbulence, this increase in velocity dispersion is largely hidden.
  We use mock X-ray observations to show that at arcmin resolution, the velocity dispersion is therefore dominated by existing large-scale turbulence and is only minimally altered by the presence of a jet. For the velocity structure of central gas uplifted by buoyantly rising lobes, we find fast, coherent outflows with low velocity dispersion. Our results highlight that 
  projected velocity distributions show complex structures which pose challenges for the interpretation of observations.
\end{abstract}

% Select between one and six entries from the list of approved keywords.
% Don't make up new ones.

%%%%%%%%%%%%%%%%%%%%%%%%%%%%%%%%%%%%%%%%%%%%%%%%%%

%%%%%%%%%%%%%%%%% BODY OF PAPER %%%%%%%%%%%%%%%%%%

\section{Introduction}

Roughly half of all known galaxy clusters show cooling times $\lesssim1\,\mathrm{Gyr}$ but lack the expected high star formation and cooling rates \citep{Fabian2012}. Active galactic nuclei (AGN) powered by supermassive black holes (SMBH) in the center of these cool-core (CC) clusters inflate buoyantly rising bubbles of hot gas, which are believed to heat the intra-cluster medium (ICM) \citep{Gitti2012,McNamara2012}. The details of the heating process are unknown. Possible mechanisms include mixing of hot bubble gas with the ICM \citep{Yang2016,Hillel2017,Hillel2018}, possibly facilitated by external cluster turbulence \citep{Bourne2019,Bourne2020}, the decay of turbulence  \citep{Zhuravleva2014a,Zhuravleva2018,Fujita2019,Mohapatra2019}, the uplift of cold gas in the wake of bubbles \citep{Guo2018,Chen2019}, the dissipation of sound waves \citep{Fabian2017,Tang2017,Bambic2019}, the dissipation of weak shocks \citep{Li2017,Martizzi2018} or the damping of cosmic ray (CR) induced Alfv\'en waves \citep{Loewenstein1991,Guo2008,Ensslin2011,Pfrommer2013,Jacob2016a,Jacob2016b,Ruszkowski2017a}. The structure of the velocity field in the ICM contains important information about the relevance of many of these processes: the amplitude and scale of the ICM turbulence can be used to infer a turbulent dissipation rate \citep{Zhuravleva2014a}, turbulent velocities combined with the cooling time set an effective range for turbulent transport \citep{Fabian2017} and the morphology of the magnetic and velocity field impacts the transport of CRs \citep{Ehlert2018}.

% RW: the transition from the first XMM measurements to the resonant scattering technique is a bit weird, especially since the result seems to be similar. How to do this better?
Observationally, many details of the velocity structure of the hot gas in the ICM remain open questions \citep{Simionescu2019}. Early results of the \textit{Suzaku} satellite found no velocity gradient within their calibration uncertainty, which provided upper limits of a few thousand km~s$^{-1}$ for cluster bulk velocities \cite[e.g., ][]{Ota2007,Ota2016,Sugawara2009,Tamura2011, Tamura2014}. Spectral analysis of \textit{XMM-Newton} data put first direct constraints on the turbulent motions in the ICM, finding large scale velocities on the order of hundreds of km~s$^{-1}$ \citep{Sanders2010,Bulbul2012,Sanders2013,Pinto2015}. Resonant scattering of lines in the core of clusters causes an apparent suppression of some lines compared to others. Thus, turbulent velocities can be constrained due to their influence on the optical depth \cite[see][for a review]{Gu2018}. 

Recent more refined analysis of resonant scattering with \textit{XMM-Newton} RGS spectra found turbulent velocities in the center of clusters on scales $\lesssim10\,\mathrm{kpc}$ of hundreds of km~s$^{-1}$ \citep[e.g.,][]{Xu2002,Werner2009,DePlaa2012,Ogorzalek2017a}. A different inference method relies on X-ray fluctuations in the smooth cluster potential that correlate with the velocity fluctuations \citep{Schuecker2004,Churazov2012,Gaspari2013a}. The method allows the computation of the velocity power spectrum, which provides constraints on the kinematic viscosity in clusters. In agreement with discussed alternative methods, recovered velocities of gas motions in the center of analyzed clusters reach a few hundred km~s$^{-1}$ \citep{Walker2015,Zhuravleva2018}. Most recently, the significant increase in spectral resolution and high broadband spectral resolution allowed the X-ray satellite \textit{Hitomi} to directly determine the bulk velocities of Perseus to be $|\vect{v}_\mathrm{bulk}|\lesssim100\ \mathrm{km}\,\mathrm{s}^{-1}$ and turbulent velocities of $\sigma_\mathrm{los}\approx 100-200\,\mathrm{km}\,\mathrm{s}^{-1}$ on spatial scales of $\approx20\,\mathrm{kpc}$ \citep{HitomiCollaboration2016, HitomiCollaboration2018}. However, detailed high-resolution velocity maps of clusters are not available and the impact of the AGN on the velocity remain unclear, which is crucial for identifying relevant heating mechanisms.

However, we can make progress by realizing that magnetic fields are tightly coupled to fluid velocities which can amplify seed magnetic fields that can themselves back-react to the flow velocities provided they are sufficiently strong. This coupling may potentially allow velocities to be deduced from magnetic fields and vice versa. And with that it provides the basis for this paper. In the framework of numerical ideal MHD, a turbulent velocity field leads to the amplification of the magnetic field $\vect{B}$, with the rate of change given by the induction equation,
\begin{equation}
\label{eq:induction}
\frac{\partial \vect{B}}{\partial t}=-\vect{v}\bdot \bnabla \vect{B}+\vect{B}\bdot \bnabla \vect{v} - \vect{B} \bnabla \bdot \vect{v}+\eta \bnabla^2 \vect{B},
\end{equation}
which relates magnetic field evolution to the velocity field $\vect{v}$ and the magnetic diffusivity $\eta$. Here and elsewhere in the paper we adopt the Gaussian cgs system of units and $\eta$ denotes the magnetic diffusivity, which has units of a diffusion coefficient. The evolution is governed by advection, stretching, compression and (numerical) dissipation of the field, respectively. 

The induction equation provides the theoretical basis for a turbulent dynamo that amplifies a seed magnetic field  \citep[e.g.,][]{Kazantsev1968,Subramanian1999,Schober2015,Beresnyak2016,Schekochihin2006}. Numerical simulations support this picture  \citep{Dolag2005,Ryu2008,Beresnyak2012,Cho2014,Roh2019}. The dynamo reaches magnetic-to-thermal pressure ratios of a few percent \citep{Schober2015,Vazza2018a}. Additional amplification is expected via compression, shocks and CRs \citep[see review by][]{Donnert2018}. AGN feedback may cause advection of galactic magnetic fields to the ICM \citep{Dubois2009,Xu2009,Donnert2009}. 

The evolution of the velocity field, in turn, is given by
\begin{equation}
\label{eq:veloevo}
\frac{\partial \vect{v}}{\partial t} = -\vect{v}\bdot \bnabla\vect{v} -\frac{\bnabla p}{\rho} - \bnabla\Phi+ \frac{1}{4\pi\rho}
\left[\vect{B}\bdot\bnabla\vect{B}-\frac{1}{2}\bnabla\bdot\left(\vect{B}^2\right)\right],
\end{equation}
where $p$ denotes the thermal pressure, $\rho$ is the gas mass density, and $\Phi$ is the (external) gravitational potential. The terms on the right describe advection, pressure force, gravity, magnetic tension and magnetic pressure, respectively. 
These equations, combined with the fact that the dynamical timescale is smaller than the lifetime of galaxy clusters, 
imply that intra-cluster medium turbulence is expected to show both, turbulent velocities as well as magnetic fields of corresponding specific energy.
For reference, equipartition of kinetic and magnetic energy density implies
\begin{align}
\frac{\rho}{10^{-25} \,\text{g cm}^{-3}} \left(\frac{|\vect{v}|}{100\,\text{km s}^{-1}}\right)^2 \approx \left(\frac{|\vect{B}|}{11.2\, \mu \text{G}}\right)^2.
\end{align}

Indeed, observations of large-scale diffuse radio structures on up to Mpc scale in galaxy clusters, i.e. radio halos, support the notion that the ICM is  magnetized \citep[see review by][]{VanWeeren2019}. Faraday RMs uncover the strength and scale of magnetic turbulence in clusters \citep{Clarke2004a}. Assuming that the magnetic field scales as a power-law with density, the expected RM from simulated fields can be compared to observations to measure its strength and injection scale. Applying this method to radio galaxies in the Coma cluster, \cite{Bonafede2010} determine a central magnetic field strength of $5\,\mu\mathrm{G}$. Additionally, \cite{Kuchar2011} analyzed the magnetic field of Hydra A and found a central magnetic field strength of $36\,\mu G$. Generally, hotter, more massive clusters show larger dispersion in RM distributions \citep{Govoni2010} and magnetic power spectra are consistent with a Kolmogorov slope \citep[e.g.,][]{Vogt2005,Guidetti2008,Vacca2012,Govoni2017}.

This paper aims to shed light on the connection between large-scale magnetic and velocity fields and their observables, i.e.\ Faraday RM and X-ray emission-line broadening, by studying simulations of decaying cluster turbulence. Simulating an AGN outburst in a Perseus-like cluster, we demonstrate that jet driven turbulence is mostly limited to the near vicinity of the jet, more specifically to the wake of the bubbles. We also relate the kinetic to the Faraday RM powerspectra and detail the velocity fields of dragged up material by the jet.

For this, we describe our initial conditions and simulation setup in Section \ref{sec:methods}. In Section \ref{sec:largescale}, we analyze our simulations of the ICM without jets. In Section \ref{sec:impactjet}, we then study the effects of AGN driven turbulence in simulations with and without pre-existing turbulence, and connect them to X-ray and RM maps. Subsequently, in Section \ref{sec:jetinduced}, we focus on AGN jet induced uplifts and conclude in Section \ref{sec:conclusion}.
 
\section{Methods}
\label{sec:methods}

We use simulations of isolated galaxy clusters to study the impact of AGN driven jets on the magnetized ICM. The equations of ideal magnetohydrodynamics (MHD) are solved on a moving mesh using the \textsc{Arepo} code \citep{Springel2010, Pakmor2016}. 
CRs created in the jets are treated as a second fluid including advection, Alfv\'enic, hadronic and leptonic losses \citep{Pfrommer2017} and anisotropic diffusion along magnetic field lines \citep{Pakmor2016a}. The simulation setup closely resembles the one in previous work \citep{Ehlert2018}, with some minor changes.

\subsection{Initial conditions}

\begin{figure*}
\centering
\includegraphics[trim=0.3cm 0.2cm 0.3cm 0.2cm,clip=true, width=\textwidth]{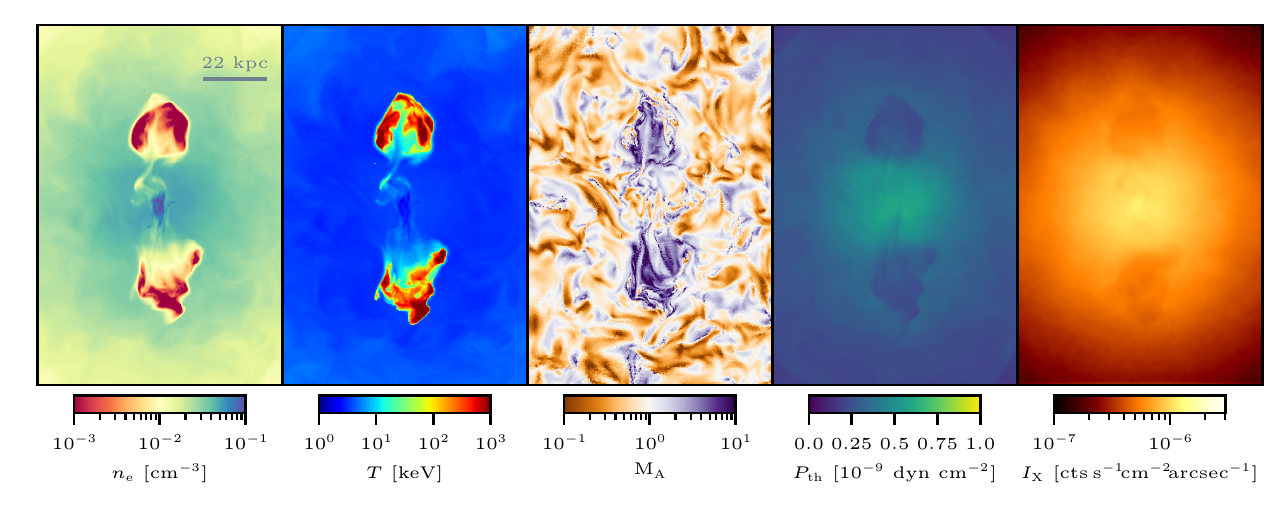}
\caption{We show thin projections ($132\,\mathrm{kpc}\times 90\,\mathrm{kpc}\times4\,\mathrm{kpc}$) of density $\rho$, temperature $T$, Alfv\'enic Machnumber $\mathrm{M}_\mathrm{A}=\sqrt{\epsilon_\mathrm{kin}/\epsilon_{B}}$, thermal pressure $P_\mathrm{th}$ and $2-12$~keV X-ray surface brightness $I_X$  centred on the BH. To reduce photon noise, the simulated X-ray exposure is $2.5$~Ms and the map has been smoothed with a Gaussian kernel of $2$~arcsec width. The bubbles are in the process of disruption by dense central gas that is accelerated upwards in the wake of the bubbles.}
    \label{fig:overview}
\end{figure*}

\begin{figure*}
\centering
\includegraphics[trim=0.18cm 0.2cm 0.2cm .12cm,clip=true, width=0.32\textwidth]{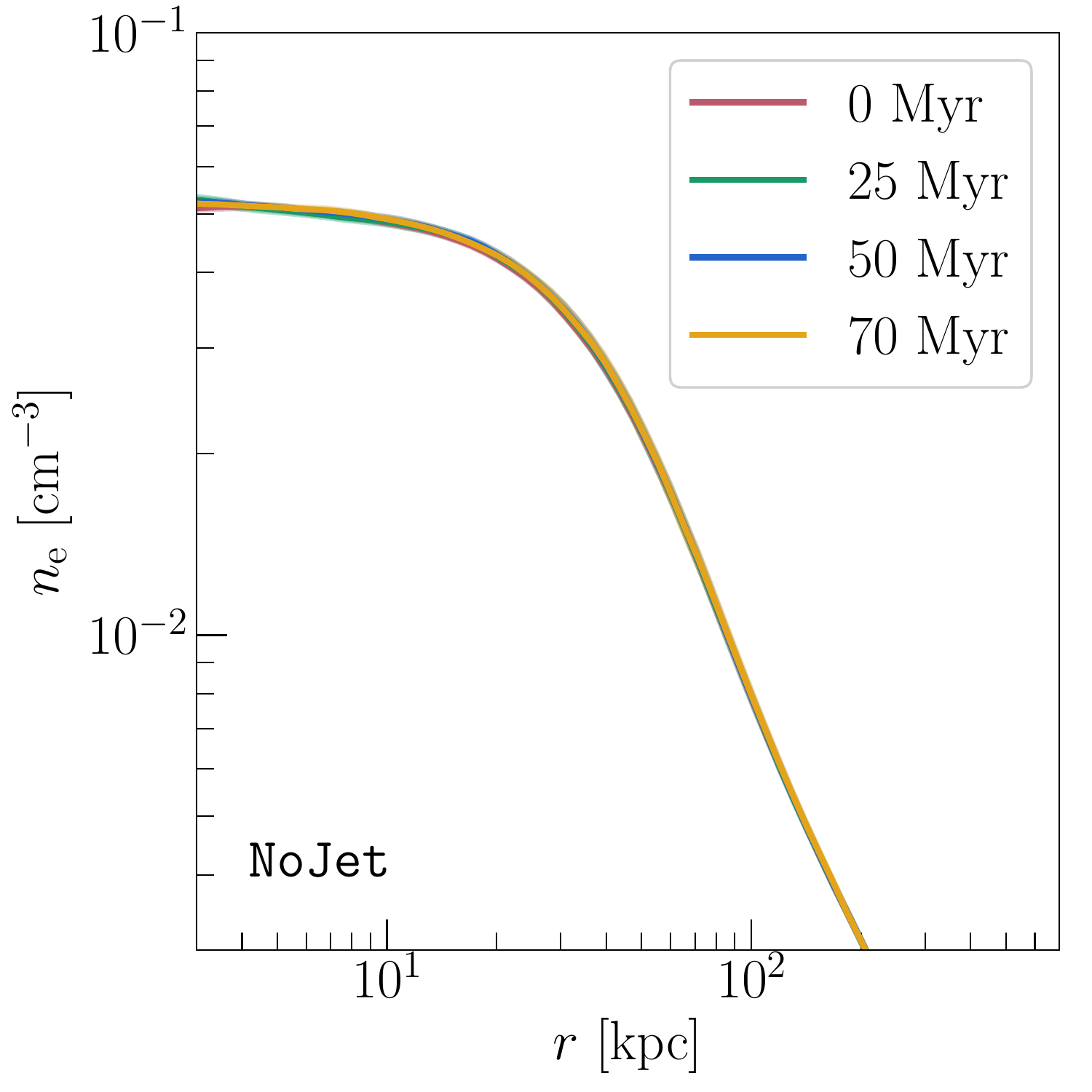}
\includegraphics[trim=0.18cm 0.2cm 0.2cm .12cm,clip=true, width=0.32\textwidth]{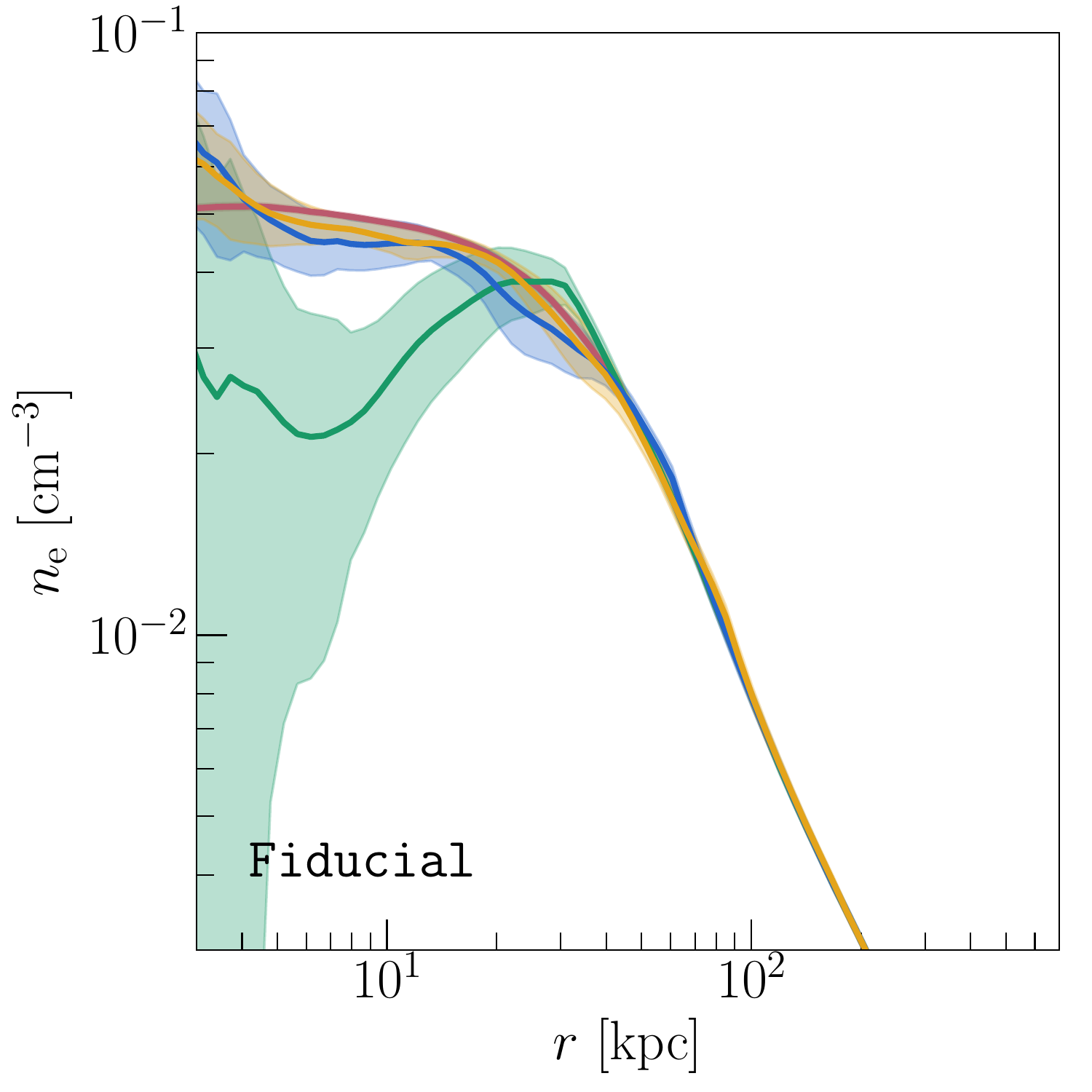}
%\hspace{0.01\textwidth}
\includegraphics[trim=0.18cm 0.2cm 0.2cm .12cm,clip=true, width=0.32\textwidth]{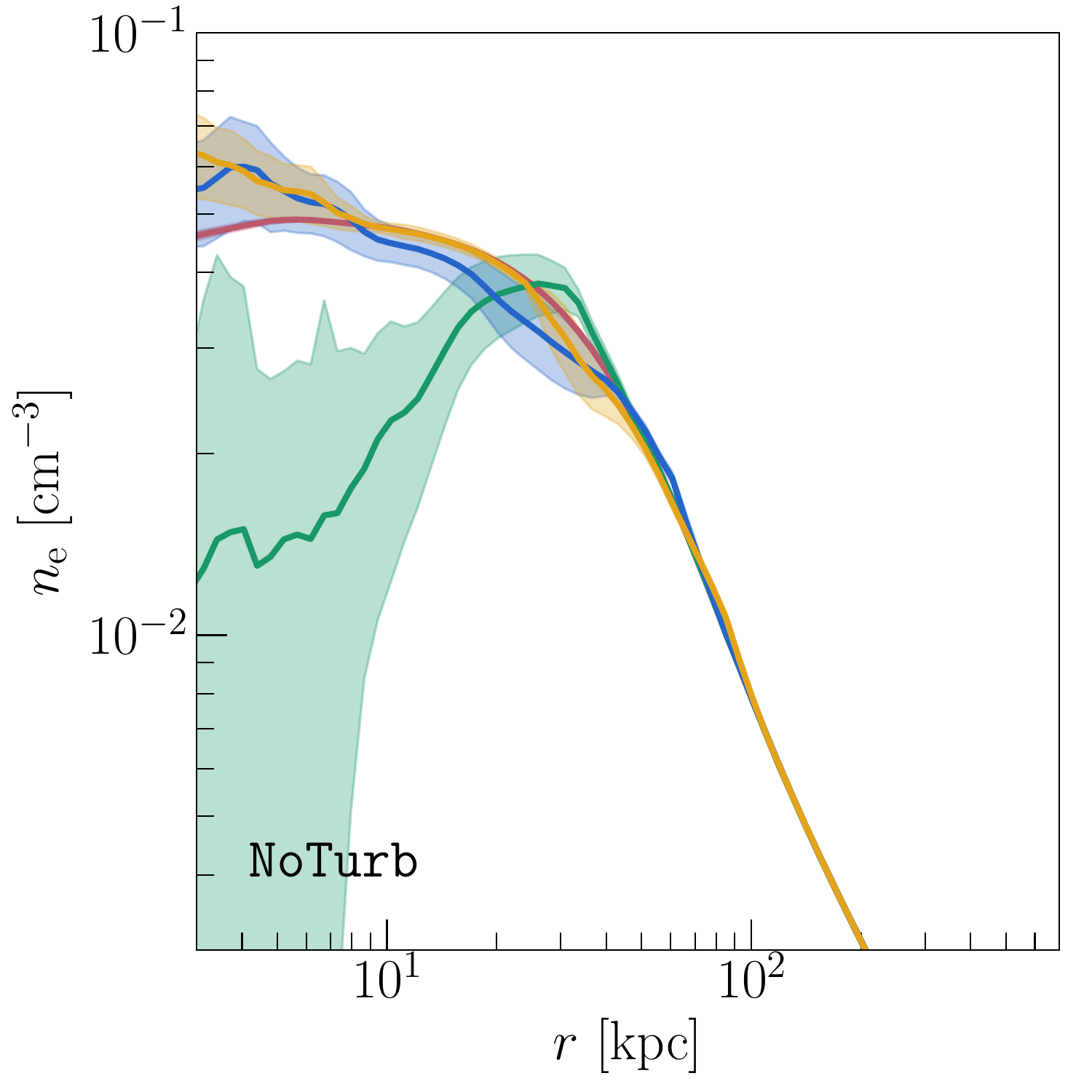}
\includegraphics[trim=0.18cm 0.2cm 0.2cm .2cm,clip=true, width=0.32\textwidth]{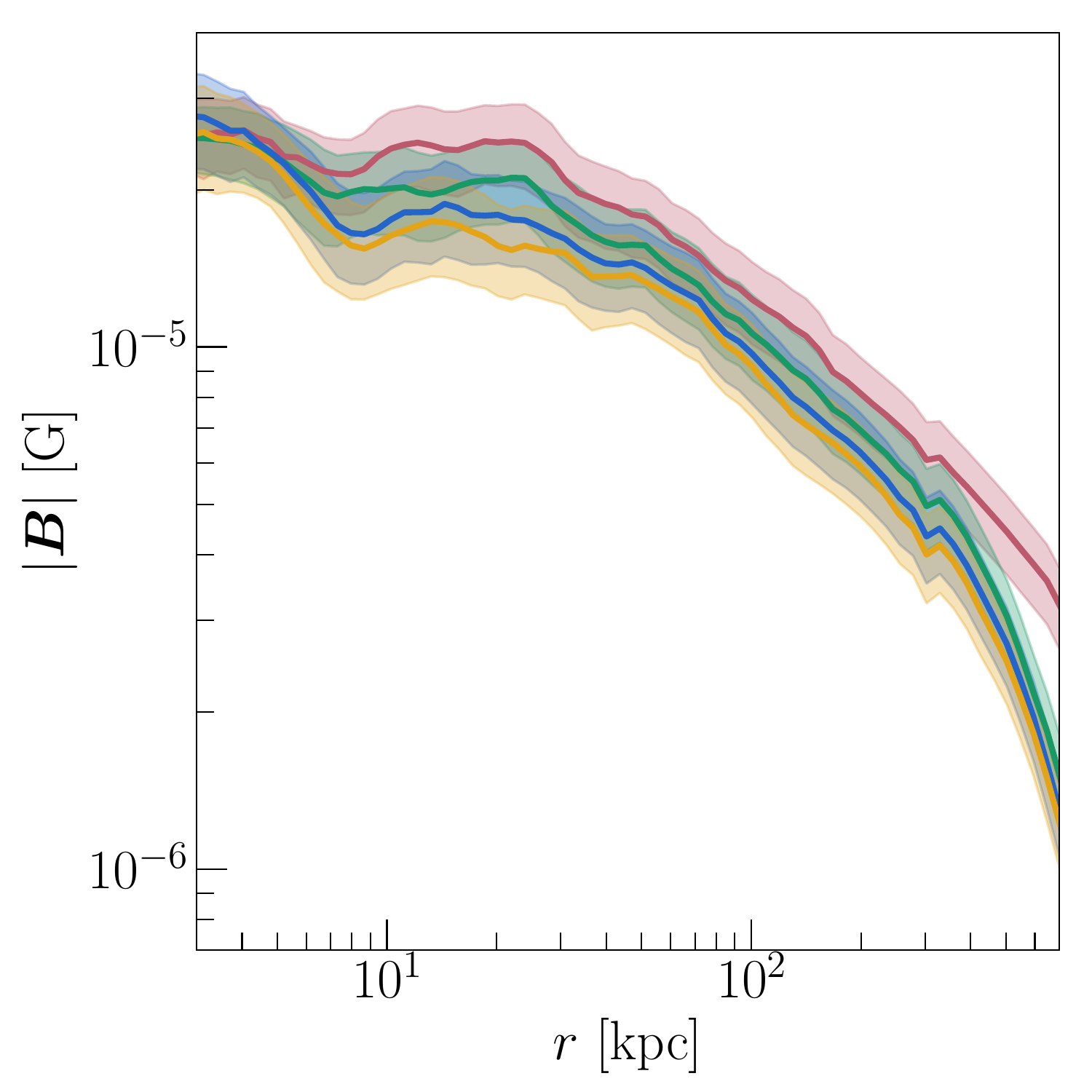}
\includegraphics[trim=0.18cm 0.2cm 0.2cm .2cm,clip=true, width=0.32\textwidth]{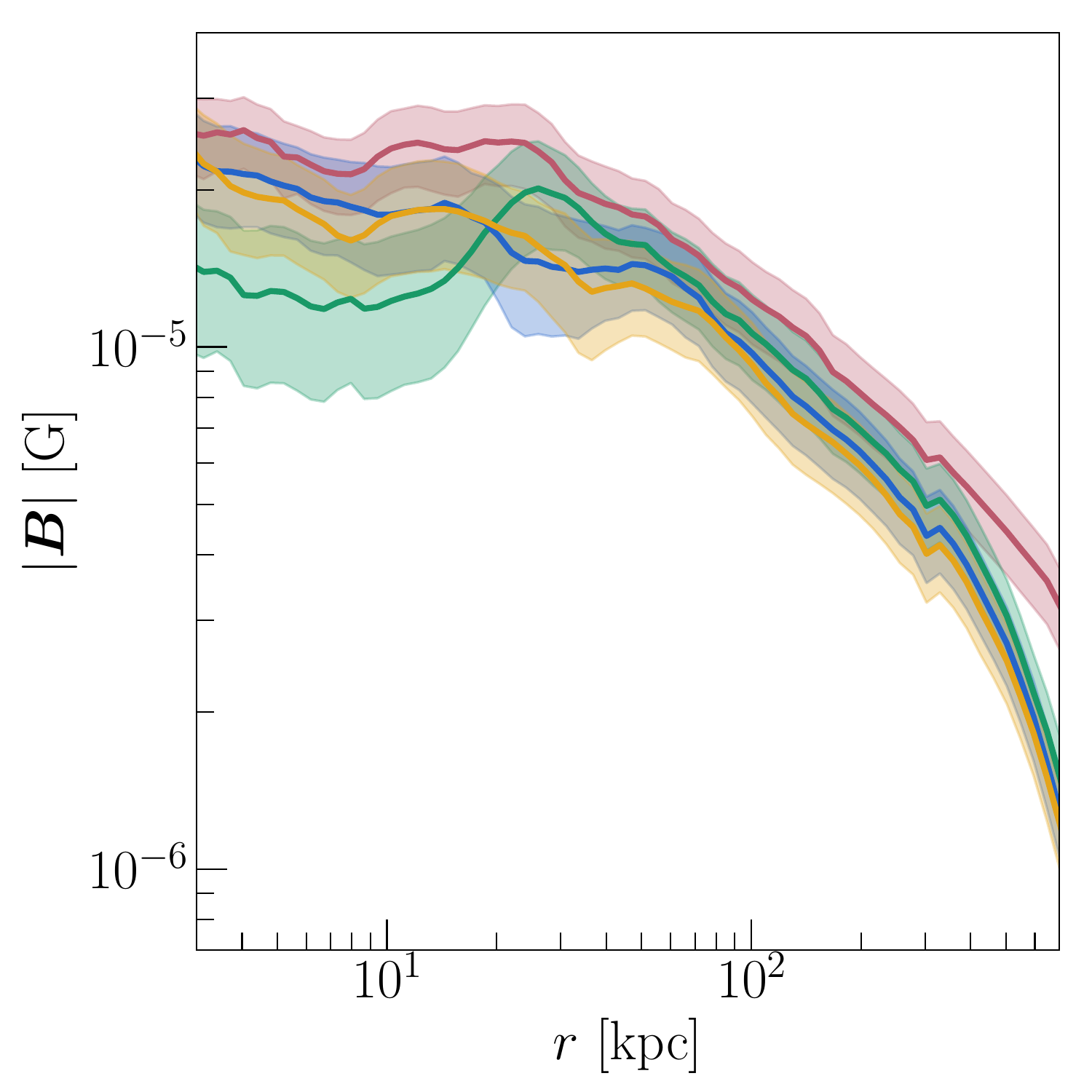}
%\hspace{0.01\textwidth}
\includegraphics[trim=0.18cm 0.2cm 0.2cm .2cm,clip=true, width=0.32\textwidth]{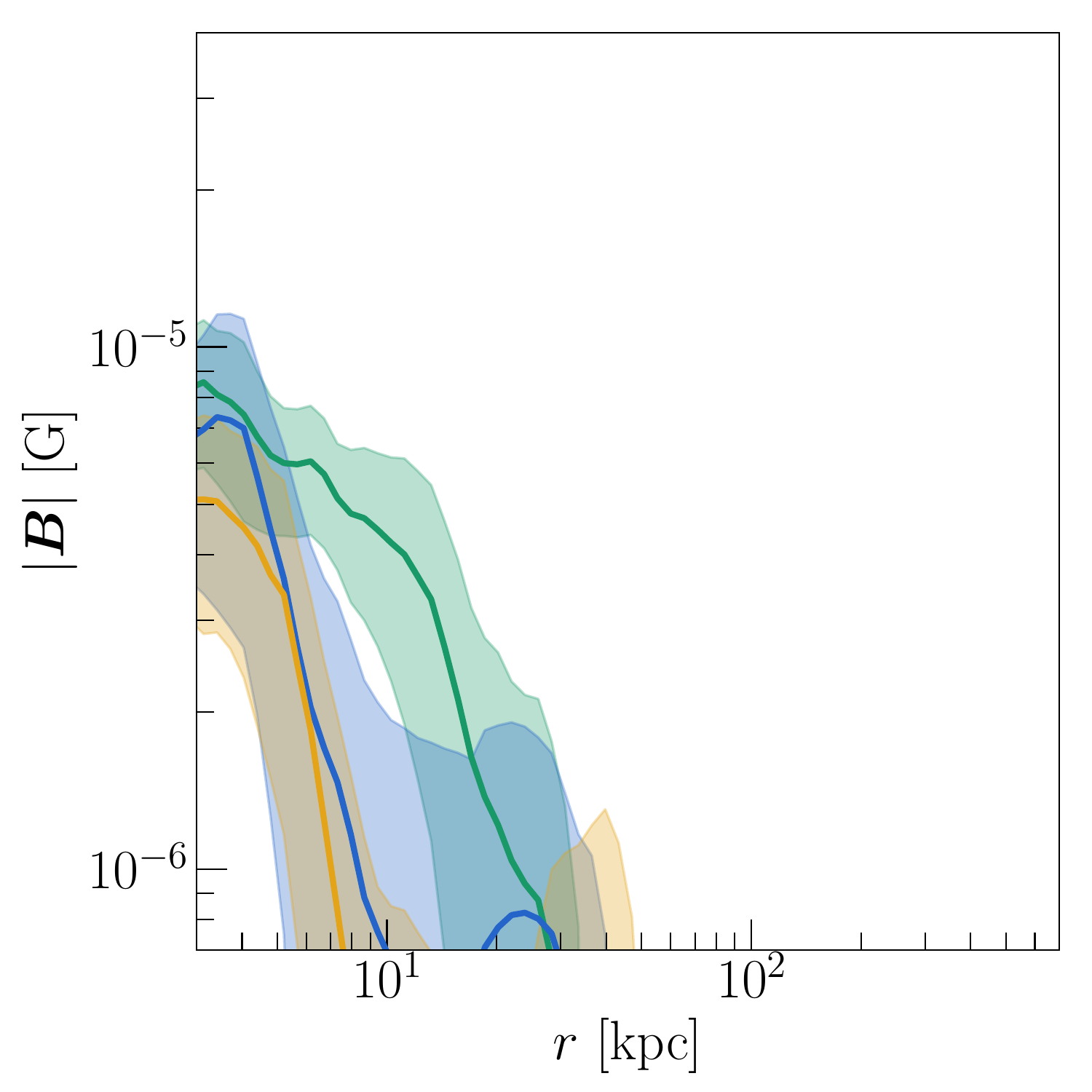}
\includegraphics[trim=0.18cm 0.2cm 0.2cm .2cm,clip=true, width=0.32\textwidth]{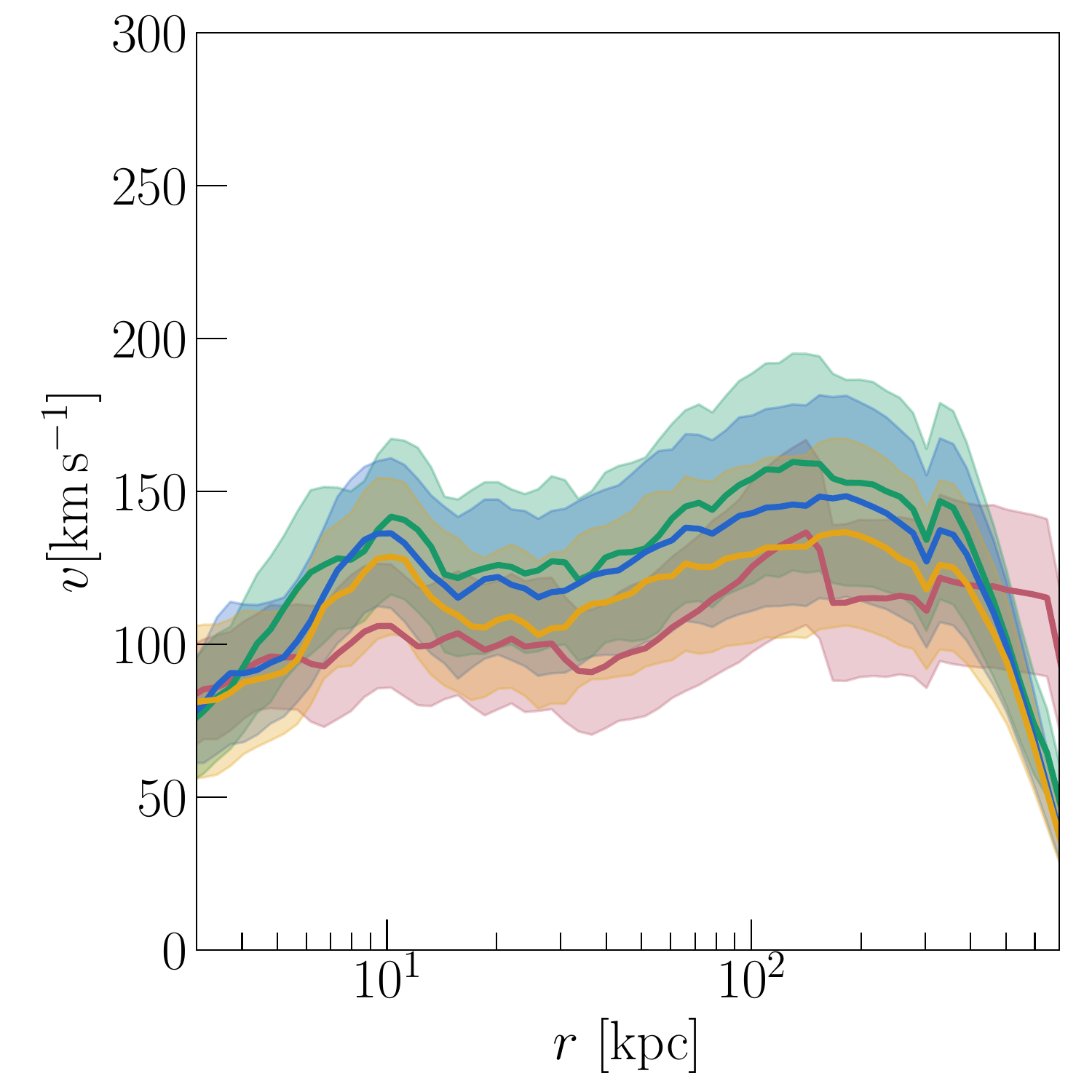}
\includegraphics[trim=0.18cm 0.2cm 0.2cm .2cm,clip=true, width=0.32\textwidth]{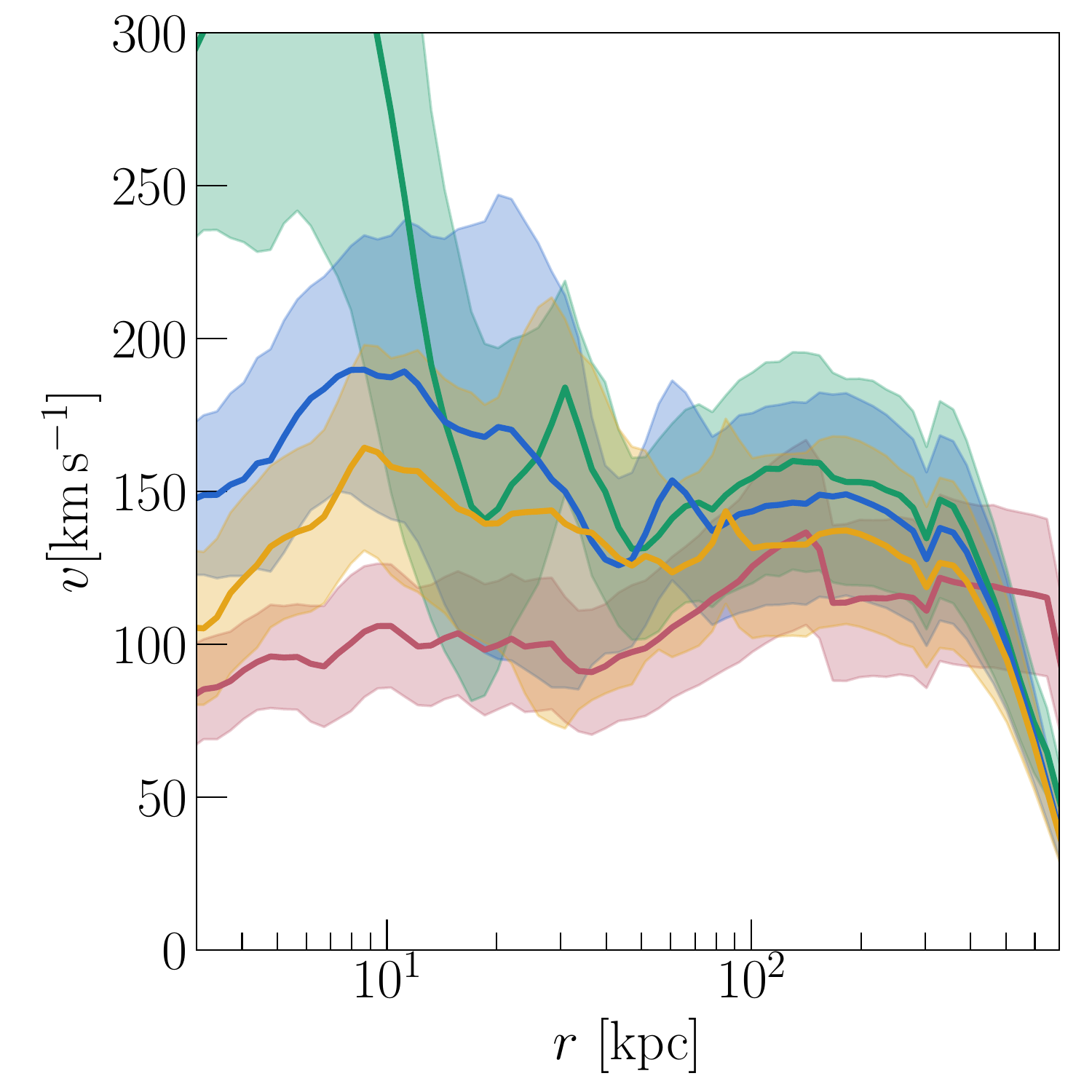}
%\hspace{0.01\textwidth}
\includegraphics[trim=0.18cm 0.2cm 0.2cm .2cm,clip=true, width=0.32\textwidth]{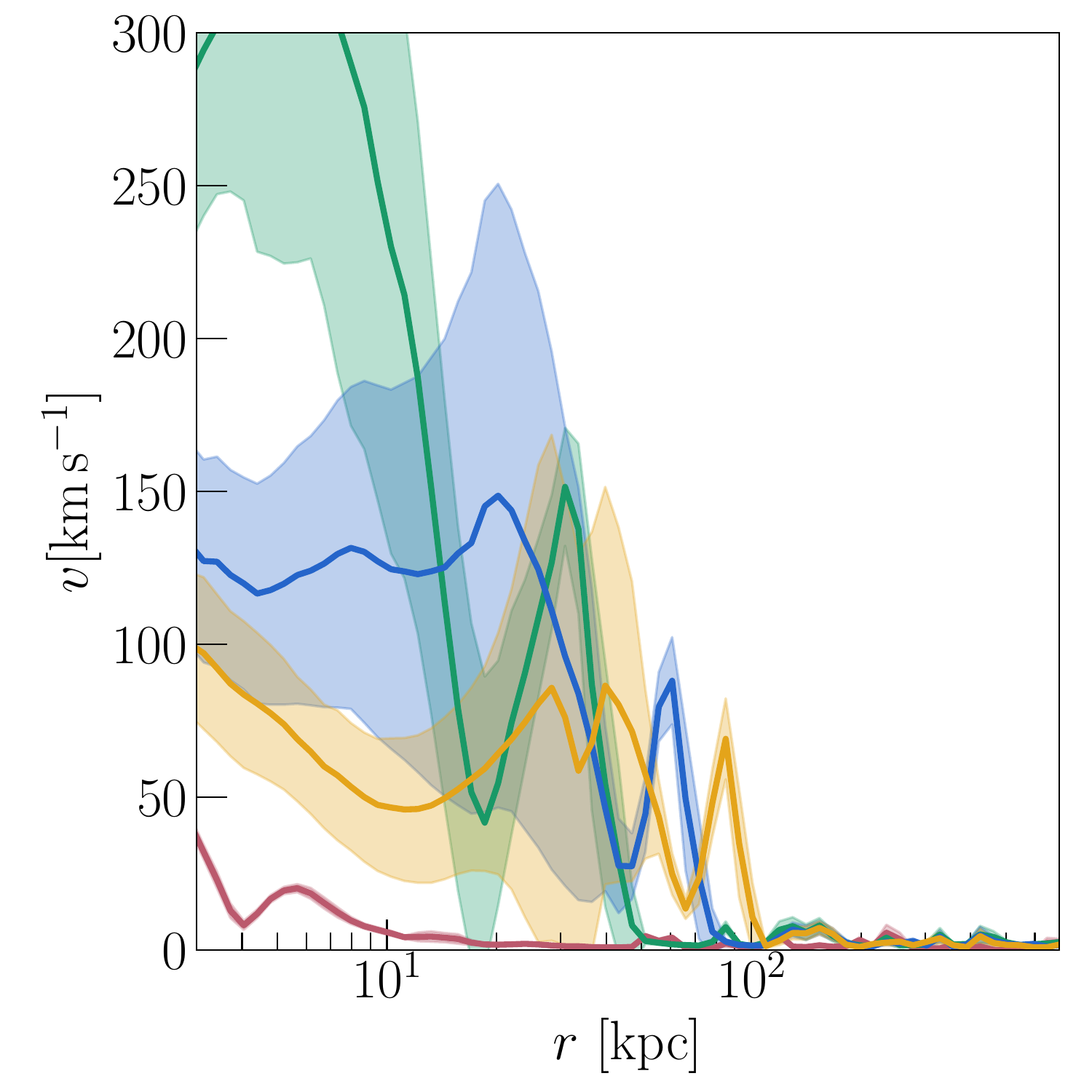}

\caption{From top to bottom we show radial profiles of volume-weighted density, square root of volume-weighted magnetic field strength and square root of mass-weighted velocities for models \texttt{NoJet}, \texttt{Fiducial} and \texttt{NoTurb} (from left to right) at stated times. The bubbles are visible as depressions at $r<20\,\mathrm{kpc}$ at $25\,\mathrm{Myr}$. The turbulent magnetic field decays as a function of time. AGN-induced velocities are limited to the vicinity of the bubble.}
    \label{fig:radialProfile_general}
\end{figure*}

\begin{figure*}
\centering
\includegraphics[trim=.1cm 0cm 0cm 0cm, clip=true, width=0.48\textwidth]{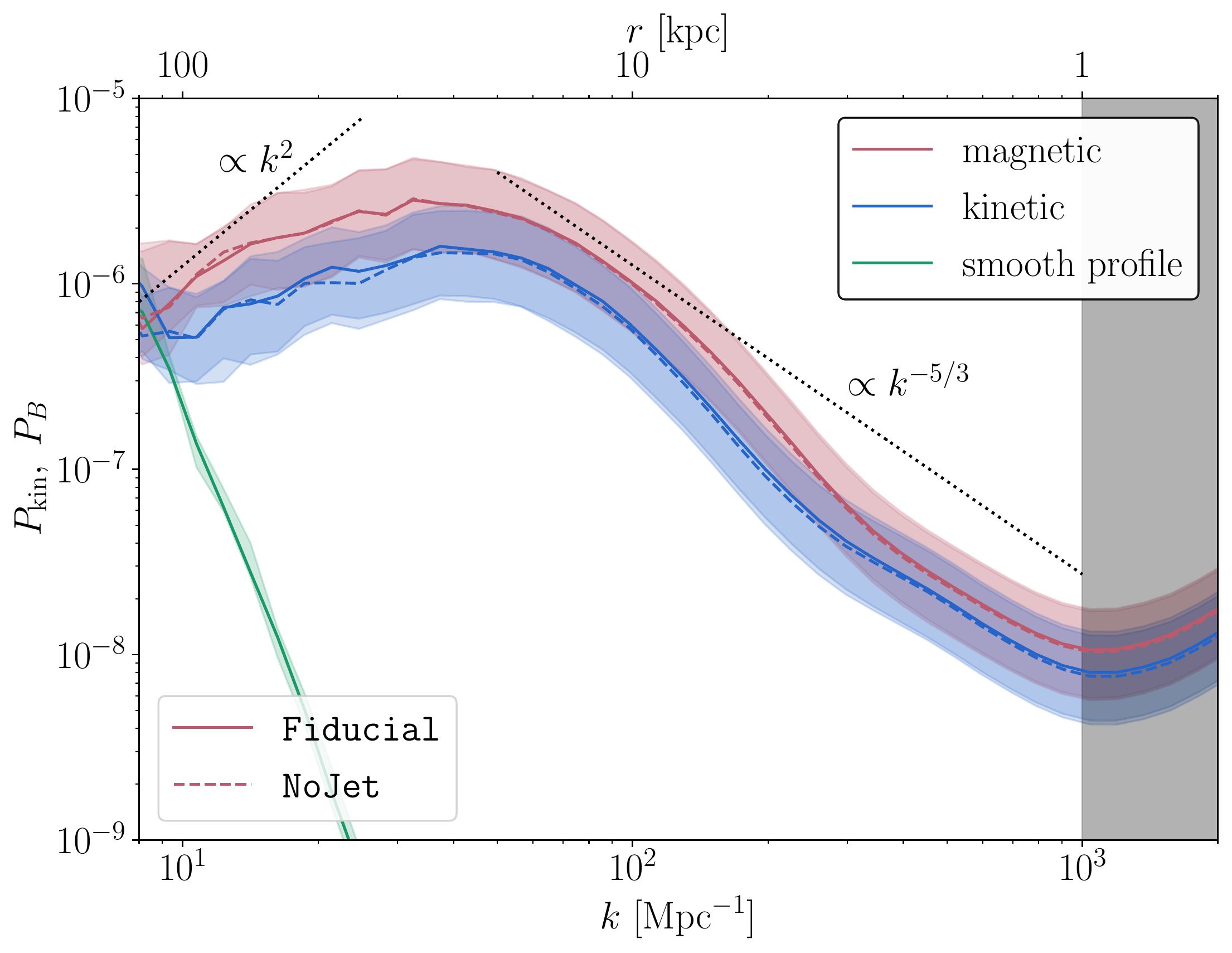}
\includegraphics[trim=0.2cm 0cm 0.2cm 0.4cm,clip=true, width=0.48\textwidth]{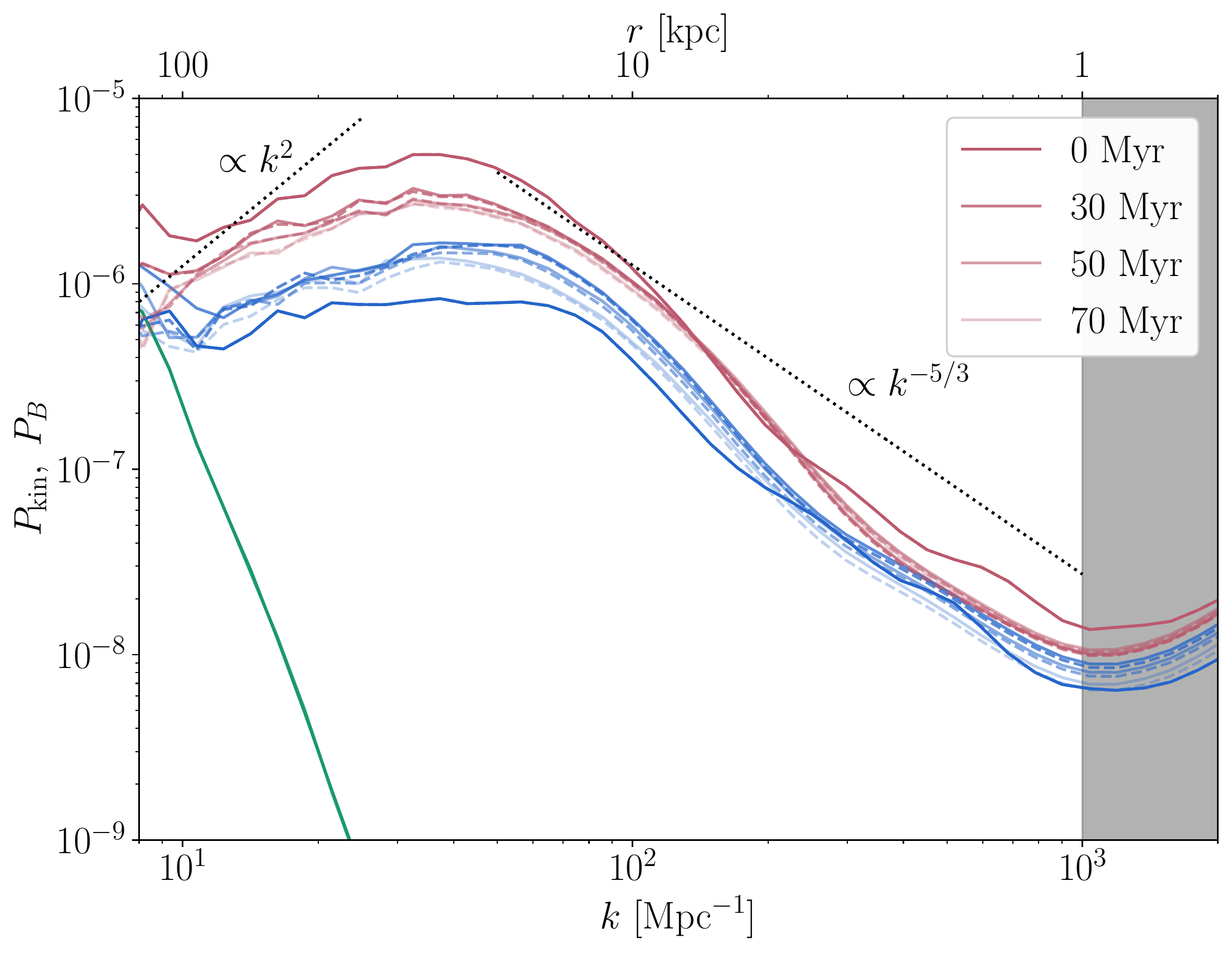}
\caption{We show one-dimensional power spectra of kinetic energy density, magnetic energy density (left panel) and its evolution at $0,\,30,\,50,\,70\,\mathrm{Myr}$ (right panel) of a BH-centred box with $L=400\,\mathrm{kpc}$ of \texttt{Fiducial} (lines) and \texttt{NoJet} (dashed). Additionally, power spectra assuming a smooth magnetic profile are plotted ('smooth profile') to quantify the contributions from the cluster profile, which only become relevant on large scales ($>200\,\mathrm{kpc}$). Shaded regions show an artificial increase in power due to leaking from higher scales which is caused by the inherent (and incorrect) assumption of periodicity. The influence of the AGN on the power spectrum is minuscule. Magnetic tension initially leads to an increase of kinetic power. After $30\,\mathrm{Myr}$ the driving by the progressively more relaxed magnetic field (as evident by decaying magnetic power) cannot keep up with the competing decay of kinetic turbulence. This leads to an overall decrease of kinetic power.}
    \label{fig:powerspecEvolution}
\end{figure*}

We model our simulated cluster after the Perseus cluster: the dark matter profile follows a Navarro-Frenk-White (NFW) profile \citep{Navarro1996,Navarro1997} with virial radius\footnote{We define the cluster virial radius as the radius at which the mean enclosed density equals $200$ times the critical density of the universe today.}  $R_{200,\mathrm{c}}=2.12\,\mathrm{Mpc}$, virial mass
$M_{200,\mathrm{c}}=8\times10^{14}\,\mathrm{M}_\odot$ \citep{Reiprich2002}, and concentration parameter $c_\mathrm{NFW}=5$. We adopt the electron number density profile from \citet{Churazov2003} rescaled to a cosmology with $h=0.67$:
\begin{equation}
\begin{split}
n_\mathrm{e} &= 46\times10^{-3}\left[1+\left(\frac{r}{60\,\text{kpc}}\right)^2\right]^{-1.8}\text{cm}^{-3}\\
 &\quad +4.7\times10^{-3}\left[1+\left(\frac{r}{210\,\text{kpc}}\right)^2\right]^{-0.87}\text{cm}^{-3}.
\end{split}
\end{equation}

We include a turbulent magnetic field in the ICM. In Fourier space the magnetic field follows a Kolmogorov spectrum on scales smaller than the injection scale $k_\mathrm{inj}^{-1}=37.5\,\mathrm{kpc}$ and white noise on larger scales. Our choice of $k_\mathrm{inj}$ is motivated by observations of RMs of CC clusters, which find magnetic fields fluctuating on scales from a few to tens of kpc \citep{Vacca2018}. Motivated by \citet{Bonafede2010}, who find their large sample of RMs for Coma to be consistent with a constant magnetic-to-thermal pressure ratio, we scale the concentric shell averaged magnetic field strength to a magnetic-to-thermal pressure ratio of $X_{B,\mathrm{ICM}}=0.05$, independent of radius. The strength of the magnetic field is motivated by radio synchrotron \citep{DeGasperin2012} and RMs \citep{Kuchar2011} by scaling magnetic field strength to the density of the Perseus cluster \citep[see Section 3.1 in][]{Pfrommer2013}. % ToDo: this last sentence needs some specifics
We refer to our previous work \citep{Ehlert2018} for the precise procedure to set up this divergence-free field.

\subsection{Jet modeling}
\label{sec:jetmodel}

The launching of the jet follows previous work \citep{Weinberger2017, Ehlert2018}, with some minor modifications: we set up two spherical \textit{injection regions} with radius $r_j=1.65\,\mathrm{kpc}$ at a distance of $3.3\,\mathrm{kpc}$ on opposite sides of a central SMBH. 
These regions serve as inflow boundary regions, in which the jet density $\rho_\mathrm{jet}=10^{-28}\,\mathrm{g}\,\mathrm{cm}^{-3}$ is kept constant throughout the injection and the thermal energy, as well as a weak toroidal magnetic field with relative energy density $X_{B,\mathrm{jet}}=0.1$ are adjusted to be in pressure equilibrium with the surrounding medium. To ensure overall mass conservation, the mass that was taken out of (or put into) the injection region is added (or removed) from the SMBH surroundings that is not within these injection regions. After accounting for the (comparably small amount of) energy required for the preparation of this state, the remaining energy is added in the form of kinetic energy creating bipolar outflows from the jet regions.

In our previous work, we assumed a jet opening angle of zero. The velocity gradient along the jet axis requires a minimum number of $\sim10$ cells to be resolved. With increasing resolution, the momentum is distributed over a smaller volume and the jet speed increases. As a result, a given amount of injected energy produced lobes at larger cluster-centric distances than observed, especially for highly resolved jets. This may hint towards a more complex small scale jet physics below the resolution limit of our simulation\footnote{Possible effects could include the interaction with a clumpy interstellar medium \citep{Mukherjee2016}, jet instabilities such as the magnetic kink instability \citep{Tchekhovskoy2016, Mukherjee2020} or an actual, wider opening angle wind component \citep{Yuan2015}.}. To cure this short-coming, here we introduce a jet opening angle, such that momentum changes are not strictly bi-directional. We determine the cone with a 30-degree half-opening angle that encloses one of the spherical injection regions and has its tip centered on the SMBH. The position vector of a cell relative to the tip of this cone is used as the direction in which the momentum is injected. We note that due to geometrical effects and the larger weighting of central cells, $30$ degree represents the largest possible angle from the jet axis that the momentum kick can be applied, while half of the momentum is injected at angles $<10$ degrees from the jet axis.

This opening angle leads to an earlier termination of the jet, producing lobes at smaller distances from the galaxy cluster center, in rough agreement with observations in the Perseus cluster (discussed in Section \ref{subsec:lobeproperties}). Figure~\ref{fig:overview} shows a snapshot of such a simulation including thin density and temperature slices, $25\,\mathrm{Myr}$ after the jet became inactive after it was on for the same time.

Since we cannot model the acceleration of CR protons self-consistently within this model, we include it in our lobes by converting thermal to CR energy until the desired CR-to-thermal energy ratio $X_\mathrm{cr,acc}=1$ is obtained while the jet is active. We magnetically isolate the injection region to inhibit unphysical diffusion of CRs. In the ICM, CRs are expected to scatter on self-generated Alfv\'en waves \citep{Kulsrud2005}. Due to inefficient Alfv\'en wave damping CRs are confined to stream down their pressure gradient $\bnabla P_\mathrm{cr}$ close to the Alfv\'en speed $v_\mathrm{A}$ \citep{Zweibel2013,Thomas2019,Thomas2020}. CRs that stream faster than the local Alfv{\'e}n speed excite the streaming instability that continuously generates Alfv\'en waves, which experience damping processes. Thereby, CR energy is effectively transformed into thermal energy via \textit{Alfv\'en heating} with power $H_\mathrm{cr}=|\vect{v}_\mathrm{A}\bdot\bnabla P_\mathrm{cr}|$. In line with \cite{Sharma2009} and \cite{Wiener2017}, we introduce an effective CR diffusion coefficient $\kappa_\mathrm{cr,A}=10^{29}\,\mathrm{cm}^2\,\mathrm{s}^{-1}$ and emulate CR streaming by including CR advection, anisotropic diffusion and account for Alfv\'enic losses \citep[see][for more details]{Ehlert2018}.

We summarize adopted parameters for our simulations in Table \ref{tab:uchangedparameters} and list the main simulations in Table \ref{table:JetPara}. 
For the high resolution runs, to focus the computational resources on the region of interest, we chose the target mass to be dependent on distance from the centre $r$,
\begin{align}
m_{\text{target},0} = 3\times10^{4}\,\text{M}_\odot \exp(r / 100\,\text{kpc}),
\end{align}
with cells at the outskirts limited to a maximum volume. In the following we focus our analysis on our high resolution simulations unless stated otherwise.

\begin{table}
\begin{center}
\begin{tabular}{ l  l  l }
   \multicolumn{3}{l}{\textbf{Jet parameters}}  \\
  \hline			
  Jet power & $P_\text{jet}$ & $10^{45}\,\mathrm{erg}\,\mathrm{s}^{-1}$ \\
  Jet life time & $\tau_\text{jet}$ & $15\,\mathrm{Myr}$ \\
  Jet density & $\rho_\text{target}$ & $10^{-28}\,\text{g}\,\text{cm}^{-3}$ \\
  Jet launching region & $r_\mathrm{j}$ & $1.65\,\text{kpc}$ \\
  CR acceleration & $X_\text{cr,acc}$ & $1$ \\
  Jet magnetization & $X_{{B},\mathrm{jet}}$ & $0.1$ \\
   & & \\
   \multicolumn{3}{l}{\textbf{Magnetic field parameters}}  \\
  \hline
  Injection scale & $k_\text{inj}$ & $37.5^{-1}\,\text{kpc}^{-1}$ \\     
   & & \\
   \multicolumn{3}{l}{\textbf{Resolution}}  \\
  \hline
  Target mass & $m_{\text{target},0}$ &  low res.: $1.5\times10^{7}\,\text{M}_\odot$ \\
  &  &  interm. res.: $1.5\times10^{6}\,\text{M}_\odot$ \\
  &  &  high res.: $4.5\times10^{4}\,\text{M}_\odot$ \\
  Jet target volume & $V_{\text{target}}^{1/3}$ & lower res.: $873\,\text{pc}$ \\
  &  & interm. res.: $405\,\text{pc}$ \\
  &  & high res.: $188\,\text{pc}$ \\
  Minimum volume & $V_\text{min}$ & $V_\text{target}/2$ 
\end{tabular}
\end{center}
\caption{Parameters for jet, the magnetic field and refinement.}
\label{tab:uchangedparameters}
\end{table}

\begin{table}
\begin{center}
\begin{tabular}{c c c c}
	\hline
   Label & Jet active & $X_{{B},\mathrm{ICM}}$ & Resolution\\
  \hline			
\texttt{Fiducial} & True & 0.05 & High \\
\texttt{NoTurb} & True & 0 & High  \\
\texttt{NoJet} & False & 0.05 & High  \\
\hline
\texttt{X25} & True & 0.25 & Interm. \\
\texttt{X5} & True & 0.05 & Interm.  \\
\texttt{X1} & True & 0.01 & Interm. \\
\end{tabular}
\end{center}
\caption{List of simulations and magnetic parameter variations. Note, run \texttt{NoTurb} is setup without magnetic fields in a hydrostatic atmosphere.}
\label{table:JetPara}
\end{table}

\subsection{Analysis}

\subsubsection{X-ray emission}

To create velocity dispersion maps, we create synthetic X-ray observations of the simulation snapshots and fit the line profile of the mock spectrum.
In particular, we employ the \textsc{pyxsim} code \citep{ZuHone2016b} using the specific internal energy and density of each gas cell, assuming a metallicity of the gas to be 0.7 solar metallicity and \citet{Asplund2009} element abundance ratios. The code makes use of the \textsc{apec} library \citep[][version $3.0.9$]{Smith2001} to calculate emission spectra in the energy range $2-12$~keV. The spectra are sampled with photon packages, taking into account thermal and Doppler broadening. The photon packages are then projected onto the detector plane, taking into account ISM absorption using the Tuebingen-Boulder absorption model \citep{Wilms2000}  assuming a column density of $4\times 10^{-20}\,\text{cm}^{-2}$, and astrophysical backgrounds are added. 
We put the galaxy cluster at a redshift of $z=0.017284$, with $1\,\mathrm{arcmin}$ corresponding to $21\,\mathrm{kpc}$ using the cosmological parameters $h = 0.67$, $\Omega_m = 0.3$, $\Omega_\Lambda = 0.7$.
We then convolve the photons with instrumental effects of XRISM Resolve using the \textsc{Soxs} library, including an Auxiliary Response File (ARF), point spread function (PSF) effects as well as energy scattering is introduced using a Redistribution Matrix File (RMF)\footnote{We use the ARF and PSF file version 20170818.}, and perform the synthetic observation with exposure time $250$ ks.

We then use the mock spectrum and fit a multi-Gaussian to the FeXXV He-$\alpha$ complex.
Similar to \citet{HitomiCollaboration2016}, we use a single width for all but the strongest line, as well as a single offset, but variable amplitudes, thus ending up with 12 fit-parameters. Note however that we omit the fitting of the weaker lines for simplicity.
We obtain the bulk velocity and line width from the fit parameters. Subtracting the square of the expected thermal broadening (assuming a single-temperature gas at $4$~keV) from the squared line width, we obtain the squared velocity dispersion.

To assess the effect of the different steps in this pipeline, we also produce ($2-12$~keV) emission weighted velocity dispersion maps as well as maps of `ideal' observations (omitting instrumental effects). We discuss these effects in Appendix~\ref{app:xray}. Note, however, that due to the idealized nature of the simulations, uncertainties resulting from temperature and metallicity variations are not captured in this setup.

\subsubsection{Faraday rotation measure}
The rotation measure $\mathrm{RM}$ is given by 
\begin{equation}
\mathrm{RM}=\frac{e^3}{2\pi m^2_\mathrm{e}c^4}\int_0^{s_\mathrm{e}} \mathrm{d}s\ n_\mathrm{e} B,
\end{equation}
where the magnetic field is integrated along the line of sight from the source at $s_\mathrm{e}$ to the observer at $s=0$.

This is done by sampling the magnetic and electron density field with a finely spaced ($\Delta x=90\,\mathrm{pc}$) 3D cartesian grid of dimensions $110\,\mathrm{kpc}-70\,\mathrm{kpc}-1.5\,\mathrm{Mpc}$, and numerically integrating along the third axis for each pixel.

\subsubsection{Kinematics of uplifted gas}

In order to study the motions of the central ICM induced by the AGN, we initialize a passive scalar within {$5\,\mathrm{kpc}$} of the SMBH to unity. This scalar is only advected with the flow and prone to dilution. To determine the velocity dispersion and mean velocity in individual pixels in the projected map we compute scalar mass weighted histograms. For this, we only consider cells with mass fractions of $>10^{-3}$. Velocity dispersion and mean velocity correspond to the variance and mean of a Gaussian that we fit to the highest peak in the velocity distribution. If fitting errors exceed $50\%$, we double the bin size in the corresponding pixel, refit the velocity distribution, and check new fitting errors. This retains  a few pixels mostly at the outskirts. We confirmed that our results remain invariant under variations of detailed parameters (bin size, mass fraction, etc.). All individual fits contain of order $10^4$ unique data points. Thereby, sufficient sampling is ensured by enforcing high resolution in the ICM. 
\newline

\section{Large-scale turbulence}
\label{sec:largescale}

To study the impact of AGN driven jets on ICM turbulence, we first discuss the evolution of the ICM in absence of jets as well as the lobe properties. Subsequently, we discuss the impact of jets in both, a quiescent and a turbulent ICM environment in Section \ref{sec:impactjet}.

First, we focus on the inherent link between equations~(\ref{eq:induction}) and (\ref{eq:veloevo}) which allows us to resort to the case of decaying MHD turbulence. Since the ICM is in hydrostatic equilibrium, i.e. the total pressure gradient is balanced by gravity, magnetic tension forces convert magnetic energy to kinetic energy as described in equation~(\ref{eq:veloevo}). Note that in this setup the magnetic field initially dominates over the kinetic turbulence. This implies that the magnetic field is not substantially replenished by dynamo processes and thus magnetic energy decreases over time, while the kinetic energy increases initially. 
This can be seen in the radial profiles of electron number density $n_\mathrm{e}$, magnetic field strength $|\bm{B}|$ and absolute velocity $v$ at different times as shown in Figure \ref{fig:radialProfile_general}. The magnetic field strength decreases independently of radius by $\approx 0.2$ dex over $70$ Myr. After $\approx 30$ Myr, stirring by magnetic tension forces becomes subdominant compared to turbulent dissipation, leading to a decrease in kinetic energy. 

To study the scale-dependence of magnetic and kinetic energy density, we show the respective one-dimensional power spectra in Figure \ref{fig:powerspecEvolution}. The magnetic power decays on most scales as shown in the right panel. To quantify the modulation of the spectrum by the profile of the cluster, we plot the power spectrum of a smooth magnetic field with magnetic-to-thermal pressure ratio $X_{B,\mathrm{ICM}}$.  Magnetic tension stirs the medium and thereby increases the kinetic power. Similarly, kinetic turbulence decreases on all scales after $30\,\mathrm{Myr}$. The modulation due to the profile of the cluster dominates the large scales of the power spectra ($k\lesssim (100\,\mathrm{kpc})^{-1}$).  On scales $(100 \,\mathrm{kpc})^{-1}\gtrsim k \gtrsim (40 \,\mathrm{kpc})^{-1}$, the magnetic field roughly follows a white noise distribution. On smaller scales a Kolmogorov slope is observed. 

\begin{figure*}
\centering
\includegraphics[trim=0.2cm .2cm 0.2cm .2cm,clip=true, width=0.7\textwidth]{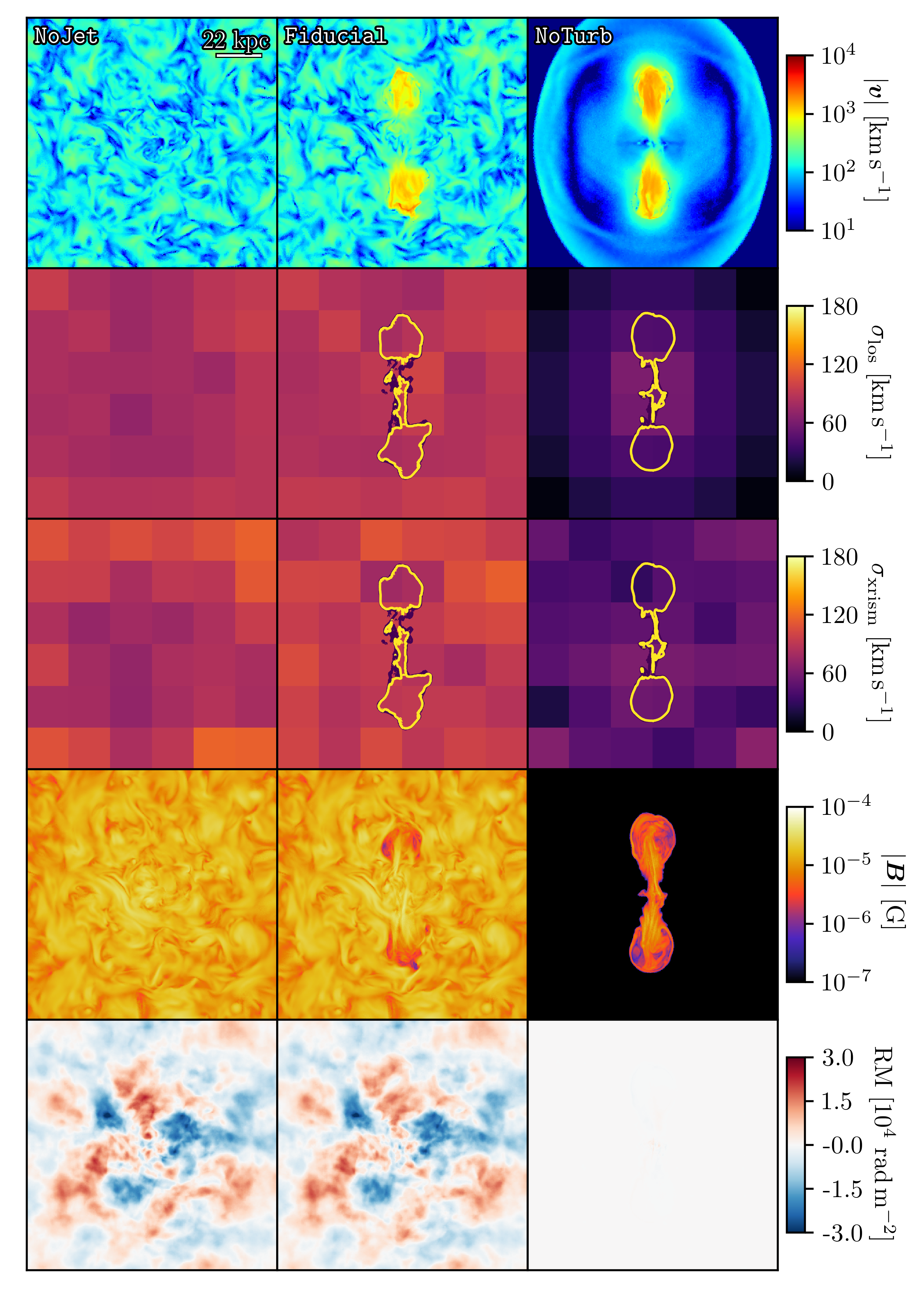}
\caption{From left to right, we display simulations \texttt{NoJet}, \texttt{Fiducial} and \texttt{NoTurb} at $50\,\mathrm{Myr}$. In the first row we show thin projections ($4\,\mathrm{kpc}$) of velocity $|\vect{v}|$ weighted with density. The second row shows velocity dispersion estimates along the line of sight $\sigma_\mathrm{los}$ directly computed from the velocity distribution in the simulation. The third row depicts the velocity dispersion obtained from fits to mock spectra $\sigma_\mathrm{xrism}$ including instrumental effects of XRISM. The fourth row shows thin projections  ($4\,\mathrm{kpc}$) of the magnetic field $|\vect{B}|$ centred on the BH  weighted with the volume. Finally, we show Faraday RM maps. Displayed images have dimensions $132\, \mathrm{kpc}\times 132\,\mathrm{kpc}$. Calculated velocity dispersion and RM encompass the full depth of the box ($1.5\,\mathrm{Mpc}$). Our initial turbulent magnetic field induces velocity dispersion at the observed level as seen for \texttt{NoJet} and \texttt{Fiducial}. The AGN affects turbulence (velocity dispersion) only in its near vicinity. Secondly, the contribution of the jet to RM is negligible.}
\label{fig:hitomi_compare}
\end{figure*}

Having established the overall behaviour, we now show synthetic observables of our simulations after $50\,\mathrm{Myr}$ when both magnetic and kinetic fields decay globally. In Figure \ref{fig:hitomi_compare}, we show (from top to bottom) slices of the velocity, the velocity dispersion along the line of sight, slices of the magnetic field and the RM of \texttt{NoJet}, \texttt{Fiducial} and \texttt{NoTurb} (from left to right) at $50\,\mathrm{Myr}$.

\begin{figure*}
\centering
\includegraphics[trim=0.2cm 0.2cm 0.2cm 0.2cm,clip=true,width = 0.9\textwidth]{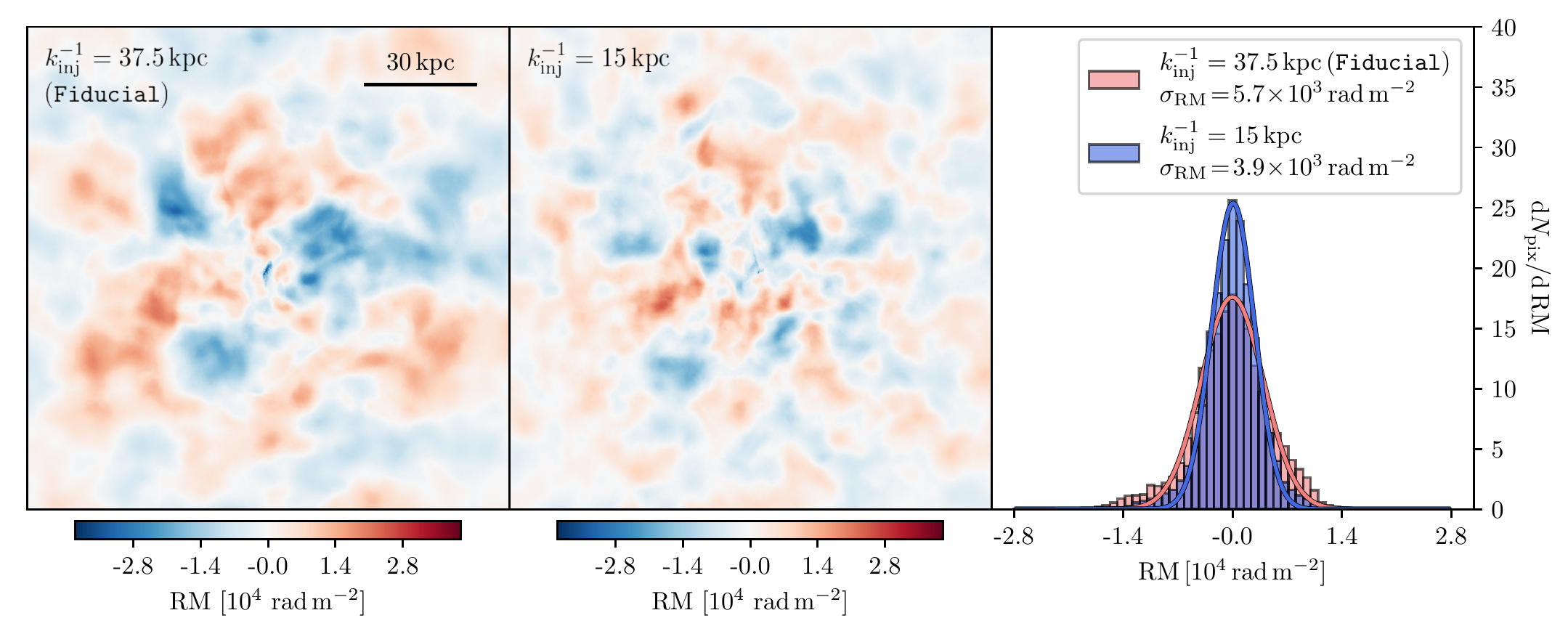}
\caption{Maps of Faraday RM of our simulations with a magnetic injection scale $k^{-1}_\mathrm{inj}=37.5\,\mathrm{kpc}$ (\texttt{Fiducial} model, left) and $k^{-1}_\mathrm{inj}=15\,\mathrm{kpc}$ (centre). Histograms of RM distribution with corresponding dispersion $\sigma_\mathrm{RM}$ are shown on the right. A decreasing injection scale implies a smaller RM dispersion.}
    \label{fig:rotation_measure_injection_scale}
\end{figure*}

We find Faraday RMs that are an order of magnitude above observed values. This is somewhat surprising as magnetic field parameters were directly taken from observations. The magnetic field of the jet has negligible effect on the overall RM (see Section \ref{sec:varymagneticfieldstrength}). We identify four effects that could be responsible for the discrepancy. (i) Possibly our adopted coherence scale is too large and we therefore underpredict depolarization. To test this hypothesis, we ran an additional simulation with an injetion scale of $k_\mathrm{inj}=15^{-1}\,\mathrm{kpc}^{-1}$, which decreases RM by $30\,\%$ (see Figure \ref{fig:rotation_measure_injection_scale}). (ii) Observed RMs are limited to few small patches in clusters that are provided by the angular extends of radio lobes. Especially the bright central radio sources in CC clusters imply a large dynamic range that challenges high-frequency polarized observations of lobes, which are negligibly affected by Faraday depolarisation. (iii) Beam smoothing artificially lowers the dispersion of observed RMs. We neglect this effect here. (iv) \cite{Vazza2018a} find their simulated magnetic fields to depart from a Gaussian distribution that is usually assumed when modeling Faraday RMs. Consequently, observations of RM possibly overestimate cluster magnetic field strengths. This highlights the relevance of cosmological MHD simulations that are able to self-consistently drive and sustain large scale magnetic fields. Moreover, this calls for a dedicated synthetic modeling of observations to take into account all possible observational effects.

The velocity dispersion $\sigma_\mathrm{los}$ corresponds to the variance of a Gaussian fit to X-ray weighted velocities along the line of sight. Gas velocities $|\vect{v}|$ in the ICM reach a few hundred $\,\mathrm{km}/\mathrm{s}$ throughout the cluster in simulations with initial turbulent magnetic field (\texttt{Fiducial} and \texttt{NoJet}). These translate to a velocity dispersion of $\sigma_\mathrm{los}\sim100\,\mathrm{km}\,\mathrm{s}^{-1}$ (second row; left and central panel), which corresponds to the observed levels by \textit{Hitomi}. This gives us another indication that our assumed field strengths are too high as this magnitude of velocity dispersion would hardly be able to sustain our initial field strengths. Varying magnetic field strengths, we find that velocity dispersion induced by fields with weaker field strengths (\texttt{X1}) yield a velocity dispersion of $\approx30-60\,\mathrm{km}\,\mathrm{s}^{-1}$ (Figure \ref{fig:hitomi_magneticvaried}). Higher field strengths (\texttt{X25}) yield a velocity dispersion of $\approx130-180\,\mathrm{km}\,\mathrm{s}^{-1}$ (see Appendix \ref{sec:varymagneticfieldstrength}). We confirmed the numerical convergence of these results (Figure \ref{fig:hitomi_resolutionvaried}). 

Thereby, the magnetic field strength in combination with the turbulent injection scale directly corresponds to a range in velocity dispersion at generally observed levels. Detailed simulations of cluster evolution therefore provide a direct link between magnetic field strength and tracers of velocity. Fundamentally, this demonstrates that RM and line broadening in X-ray spectra are just two sides of the same coin with their physical origin in cluster turbulence. The presence of magnetic turbulence implies velocity dispersion and vice versa. Our simulations of decaying turbulence are meant as a proof of concept to explore the relationship between velocity fields and magnetic fields. In reality, perturbations to the cluster potential induced by merger or accretion induce velocity fluctuations that excite a small-scale dynamo, which leads to exponential amplification of the magnetic field on the dissipation scale. After equilibrating with the kinetic energy at this scale the magnetic field might inverse cascade and is expected to saturate at a fraction of the kinetic energy \citep[e.g.,][]{Schober2015,Dominguez-Fernandez2019}, unlike what we have adopted in the initial conditions. Our results highlight the necessity for a self-consistent turbulent dynamo to amplify the magnetic field to equilibrium levels to reliably determine the normalization in the relationship between RM and ICM velocity dispersion. We conclude that one can predict levels of turbulence and magnetic field strengths by observing the other, provided the turbulent injection scale and the growth history is correctly simulated. 

\begin{figure*}
\centering
\includegraphics[trim=0.2cm 0.2cm 0.2cm 0.2cm,clip=true,width = 0.9\textwidth]{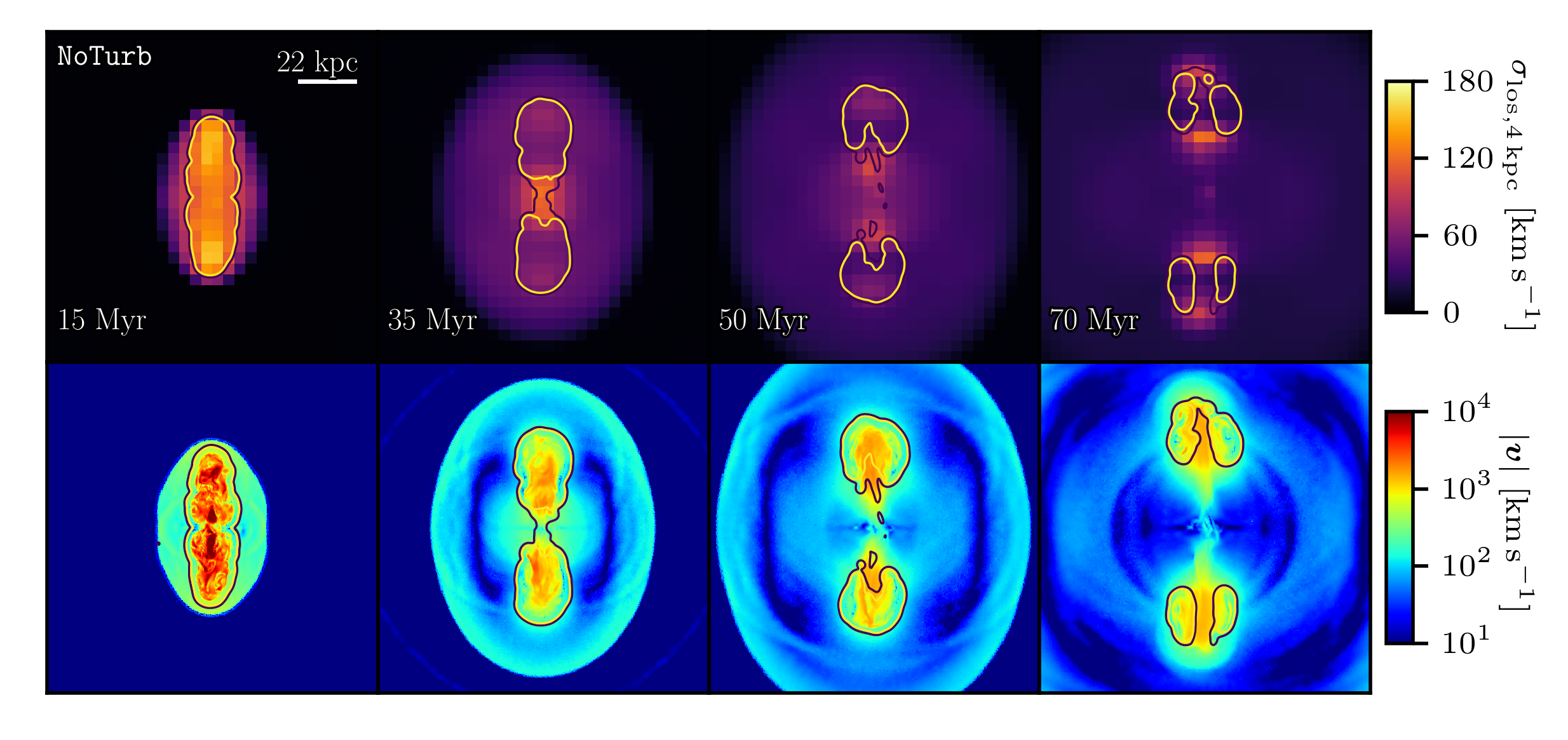}
\caption{We show the velocity dispersion of the ICM (top) and slices of the velocity (bottom) at different times (left to right) of the model $\mathtt{Noturb}$. We overlay contours of slices of the jet tracers $X_\mathrm{jet}=\left\{10^{-3},\,10^{-2}\right\}$ (purple and yellow) to highlight the location of the lobes. Displayed images have dimensions $132\, \mathrm{kpc}\times 132\,\mathrm{kpc}$. After passing of the initial shock wave, the rising bubbles only causes an increased velocity field in the immendiate wake of the bubbles.}
    \label{fig:velocity_impact_jet}
\end{figure*}

\begin{figure*}
\centering
\includegraphics[trim=2cm 1cm 0.2cm 0.8cm,clip=true,width = 0.7\textwidth]{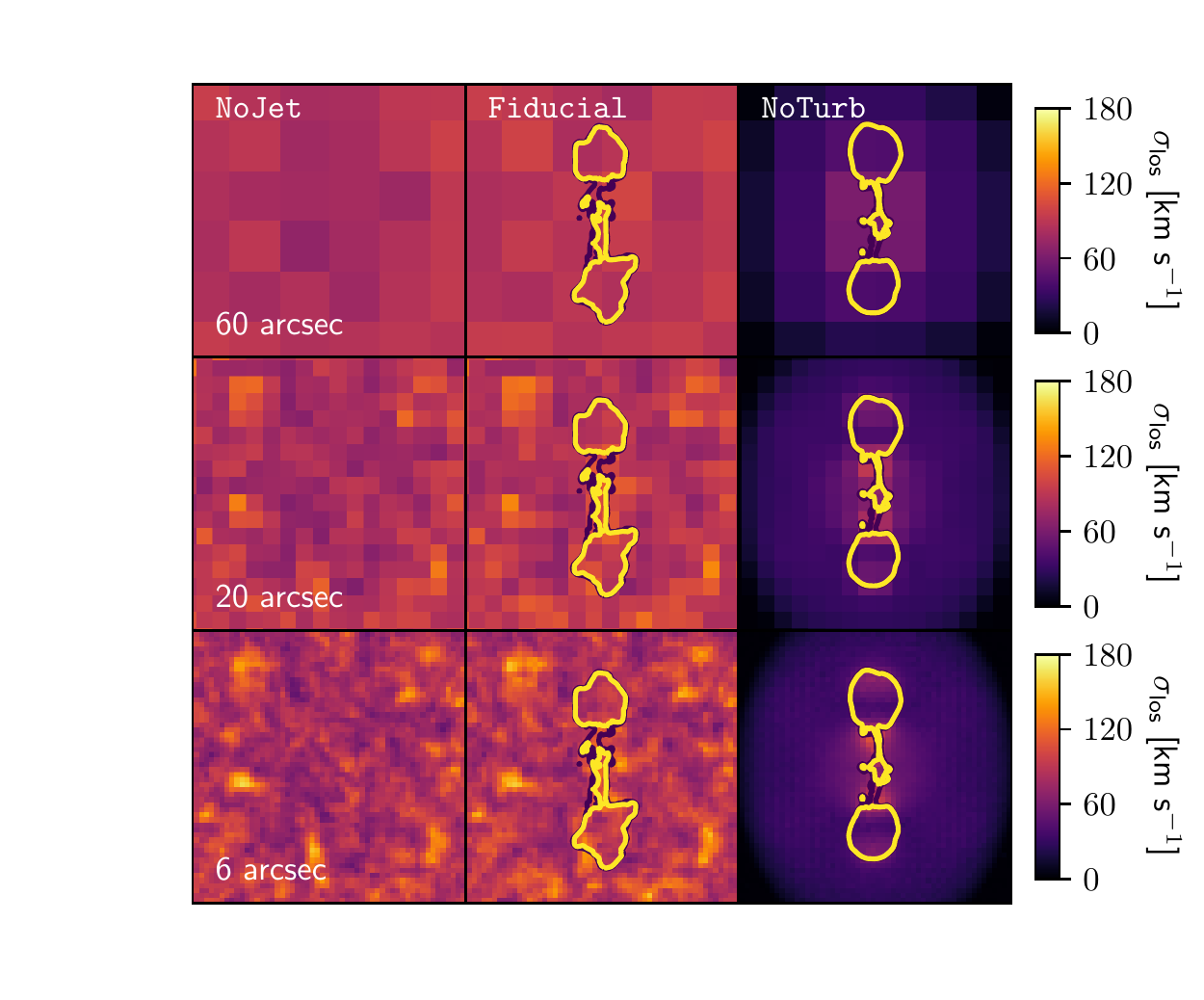}
\caption{Emission weighted velocity dispersion maps with different spatial resolution.}
    \label{fig:resolution_dependence}
\end{figure*}

\section{Impact of jets on the ICM}
\label{sec:impactjet}
\subsection{Lobe properties}
\label{subsec:lobeproperties}

In Figure \ref{fig:overview}, we show the bubble after $50\,\mathrm{Myr}$, i.e. $25\,\mathrm{Myr}$ after the jet became inactive. From left to right, we show density, temperature, Alfv\'enic Mach number, thermal pressure and X-ray emissivity. The bubbles appear as cavities in the integrated X-ray emissivity. The Alfv\'enic Mach number corresponds to the ratio of kinetic and magnetic energy density, which, in parts of the ICM reaches $M_\mathrm{A}<1$. Thus the magnetic pressure is dominant and stirs the medium. The jet inflated lobes that rise buoyantly in the cluster atmosphere. At this time, the Rayleigh-Taylor instability starts to markedly disrupt the bubbles. In the wake of the bubbles dense gas is pulled upwards from below. The core of both bubbles is already filled with denser ICM and the southern bubble is in the process of splitting (see density and temperature map). The thermalization of kinetic energy during jet propagation and lobe formation drastically lowers the magnetic-to-thermal pressure ratio compared to the initial value in the jet, $X_{B,\mathrm{jet}}$. Consequently, the magnetic field becomes subdominant in the lobes (see Alfv\'enic Mach number $\mathrm{M}_\mathrm{A}>1$). In the thermal pressure map the bubbles appear as cavities due to the significant pressure support by CRs, as is evident from the large CR-to-thermal pressure ratio $X_\mathrm{cr}\geq0.5$. 

% Comparison to Perseus bubbles (dimensions, shape)
The total work required to inflate the lobes can be quantified as the sum of $p_{\text{tot}}V$ of all cells with a jet scalar contribution $X>10^{-3}$ (assuming that the bubbles are in pressure equilibrium with the ambient ICM). Here $p_{\text{tot},i}$ is the total pressure of cell $i$, taking into account thermal, CR and magnetic pressure and $V$ its volume. In the present snapshot the energy determined this way is $\sum_i p_{\text{tot},i}V_i = 9.9 \times 10^{58}$ erg for both lobes combined. The energy estimate changes by roughly $30 \%$ if the threshold for $X$ is changed by an order of magnitude, highlighting the robustness of this result.

The thermal, CR and magnetic energies in the lobe are $1.1 \times 10^{59}$ erg, $7.6 \times 10^{58}$ erg and $7.2 \times 10^{56}$ erg, respectively, constituting a total lobe enthalpy of $2.8 \times 10^{59}\,\text{erg} = 2.87\, \sum_i p_{\text{tot},i}V_i$ (the total injected energy is $4.7\times 10^{59}$). We note that CR and thermal energy in the active jet are brought to equipartition by construction, implying an enthalpy of $3 \, \sum_i p_{\text{tot},i}V_i$ at these times. The subsequent dominance of the thermal component is due to CRs diffusing out of the lobe during the $35$~Myr since the termination of the jet. Adiabatic expansion of the lobe drives the lobe towards being more CR dominated (since the adiabatic index of the CR fluid is lower). However, this effect is evidently smaller than diffusive CR losses. We can also infer that about 60 per cent of the injected jet energy is no longer associated with the lobes. This energy has been (i) dissipated in the bow shock during jet propagation, (ii) invested in $p\rmn{d}V$ work on the ambient ICM during the formation and expansion of the lobes, and (iii) lost by diffusing CRs that have escaped the lobes as well as hydrodynamic mixing of lobe material with the surroundings.

The extent and shape of the lobes can be evaluated from the X-ray emissivity image, with a lobe diameter of about $20\,\mathrm{kpc}$ at a distance of $25\,\mathrm{kpc}$ from the center. 
These values can be compared to literature values of the lobes in the Perseus Cluster, e.g. \citet{Birzan2004,Rafferty2006,Diehl2008}, which have similar dimensions and positions.
We are therefore confident that the behaviour of these lobes is similar enough to meaningfully compare their effect in the simulated ICM to observational data.

\subsection{Dynamical impact of jets}
\label{sec:magneticvelocityconnection}

Having presented the properties of the ICM and the lobes, we move on to focus on the effect of a single jet outburst on observable ICM properties. 

The inflation of the bubble is visible as a drop in both density and magnetic field strength within the inner $10\,\mathrm{kpc}$ at $25\,\mathrm{Myr}$ in Figure~\ref{fig:radialProfile_general}. The compression of the upstream gas due to the shock front is also evident in the profile shown in Figure~\ref{fig:radialProfile_general}. From the magnetic field and particularly the absolute velocity profiles, it becomes evident that the jet does have some impact on the kinematic structure of the gas within the range of the lobe itself, however not beyond this radius. The simulations without initial turbulence clearly show the limited range of the turbulent driving of the AGN lobes. 

Turning to the spatial distribution of the turbulence using line of sight velocity and velocity dispersion in Figure~\ref{fig:hitomi_compare}, it becomes evident that the arcmin spatial resolution mostly reveals large-scale turbulence which in our simulations originates from the initial conditions, rather than the jet. However, a small, but subdominant effect from the jet is observable, yet only differentially. In the case of a quiescent ICM it becomes particularly clear that the impact on the ICM turbulence is only present at the location of the lobes and in their past trajectory, disfavoring the idea that the lobes can sustain ICM turbulence throughout the entire cluster core and on scales up to several tens of kpc. This also suggests that the observations probing these spatial scales might not directly probe AGN induced turbulence, but rather turbulence induced by other processes such as fluctuations of the cluster potential as a result of mergers or accretion, substructure infall, or thermal instability (all of which are not included in these simulations). We discuss the required spatial resolution to detect the lobe-induced turbulence in the next subsection.

Although our analysis is limited to a single jet outburst, Figure~\ref{fig:velocity_impact_jet} illustrates that  jet-driven turbulence is spatially constrained to the wake of the rising bubbles. We focus on our model $\mathtt{NoTurb}$ and show the velocity dispersion of the ICM (top) and slices of the absolute velocity (bottom panels). The initial AGN outburst drives a shock wave into the ICM, which induces small-amplitude vorticity, which quickly decays. Significant turbulence is therefore limited to the wake region of the jet/bubbles, which is then advected upwards as those buoyantly rise. Our jet parameters are modelled after detected bubbles in Perseus. While more powerful AGN events are expected to posses larger wake regions, they are not observed in Perseus. In addition, duty cycles are not observed in Perseus to be short enough for a concatenation of small outbursts to be able to sustain turbulence in the entire core region. Nevertheless, the uncertain degree of jet precession may potentially increase the level of isotropy of the jet driven turbulence somewhat.

\cite{Zhuravleva2014a} infer the ICM heating rate due to dissipation of turbulence from X-ray observations of CC clusters. They conclude that the turbulent heating rate is sufficient to halt cooling in these clusters and suggest it as the main heating mechanism \citep{Zhuravleva2016a}. In contrast, \citet{Mohapatra2019} find the required level of turbulence for turbulent dissipation to be the dominant heating mechanism to be inconsistent with \textit{Hitomi} measurements. Our results indicate that lobes of the size of the ones in Perseus cannot be the sole driver for this level of turbulence. This is in agreement with findings of other simulations in the literature \citep{Reynolds2015, Bourne2017, Bambic2018, Bambic2019}.
Interestingly, simulations in self-regulated setups \citep[e.g.][]{Lau2017, Li2017} show slightly higher levels of turbulence.

\subsection{Observational effects}

% full mock vs vel. disp map
In Figure \ref{fig:hitomi_compare} we show emission weighted velocity dispersion maps (second row) and  velocity dispersion maps determined by fitting spectra of synthetic XRISM observations (third row). Overall, the two maps show agreement, highlighting the robustness of the method, however with a noteworthy systematically higher velocity dispersion in the outer regions of the synthetic observations. To understand this discrepancy, it is important to realize that the fitted line broadening is a superposition of thermal and Doppler broadening. Thus, to infer the velocity dispersion, we subtract the effect of thermal broadening, for which we assume a temperature of $4$~keV. Since the cluster is set up as a cool-core cluster, its real temperature in the center is slightly lower in comparison to the edges of the projection, which leads to an incomplete subtraction of thermal broadening at the edges and thus an overestimate of the velocity dispersion in these regions. 
Note that the temperature differences in our simplified setup are likely lower than in reality since we do not include cooling or cosmological environment in our setup.
In Appendix~\ref{app:xray} we disentangle the instrumental effects from the effect of fitting the spectra on the inferred map. 

In addition, we show emission weighted velocity dispersion maps at different spatial resolutions in Figure~\ref{fig:resolution_dependence}. While the $60$~arcsec resolution map shows the overall level of velocity dispersion, the $20$~arcsec resolution map already resolves the largest modes of the turbulence. Most strikingly on display at the highest resolution map ($6$~arcsec) is the fact that the \texttt{Fiducial} map seems to be a superposition of the external turbulence of the \texttt{NoJet} and the jet driven turbulence of the \texttt{NoTurb} runs. This indicates that the degree of ICM turbulence into which the jets are launched does not impact the extent to which they drive more turbulence. We speculate that this breaks down once the shape and position of the jet inflated lobes is substantially altered by existing ICM turbulence. 

In summary, we find that the kinematic impact of the lobes is very localized, which is only identifiable at high resolution, when the lobes are spatially resolved. This is particularly true when considering the bulk velocities with superimposed lobe velocities seen at an inclination (not shown here). We note that we did not create full synthetic observations (in Figure~\ref{fig:resolution_dependence}) and ignored the effect of photon noise in this example. However, Figure~\ref{fig:resolution_dependence} illustrates the wealth of additional information higher resolution velocity dispersion maps contain about the state of turbulence in the ICM and its origins.

We conclude that AGNs only drive turbulence locally. This is in line with previous work, which find AGNs to be very inefficient drivers of kinetic energy \citep{Reynolds2015,Yang2016,Bourne2017,Hillel2017a,Prasad2018}. Consequently, cluster turbulence seen by \citet{HitomiCollaboration2016} appears rather related to sloshing motions initiated by mergers \citep{ZuHone2018,Walker2018a} and/or cosmological flows. Having shown that the main impact on gas flows from jets is in the wake of inflated lobes, we focus on this region in the following.

\section{Jet-induced uplift of the ICM}
\label{sec:jetinduced}

Cold gas is observed as disks and/or filamentary structures in the center of galaxy clusters \citep[e.g.,][]{Koekemoer1999,Salome2006,Russell2019}. Submilimeter to optical observations reveal the complex velocity structures of cold gas. While disks show circular motions, filaments often surround bubbles or are found in the downstream regions. Predominantly smooth velocity gradients in filaments provide additional evidence that AGNs have a strong influence on their velocity structures \citep[e.g.,][]{Werner2014a, Tremblay2018,Gendron-Marsolais2018}. Its origin is under debate. One idea is that cold gas condenses out of the hot phase when the ratio of cooling time $t_\mathrm{cool}$ and free-fall time $t_\mathrm{ff}$ fall below a critical value $t_\mathrm{cool}/t_\mathrm{ff}<10$. This cold gas is then uplifted by an AGN \citep{Gaspari2012,Voit2017}. Alternatively, the thermal instability may be triggered by the turbulent uplift by an AGN \citep{McNamara2016,Olivares2019,Martz2020} or possibly a sloshing galaxy \citep{Vantyghem2019}. Simulations by \cite{Beckmann2019a} find evidence for the former two.

In the following, we focus on the flow patterns of gas that is dragged up by the jet from the vicinity of the SMBH. We employ our idealized, non-radiative MHD setup to separate motions induced by cooling from motions caused primarily by the AGN. Motions induced by condensating material that is collected in the so-called cloud-growth regime \citep{Gronke2018,Li_Hopkins2020,Sparre2020} require higher resolutions and are usually studied in dedicated simulations of cooling clouds submerged in hot winds. Consequently, this effect is not present in our setup.

\begin{figure*}
\centering
\includegraphics[trim=0.2cm 0.2cm 0.2cm 0.2cm,clip=true, width=0.76\textwidth]{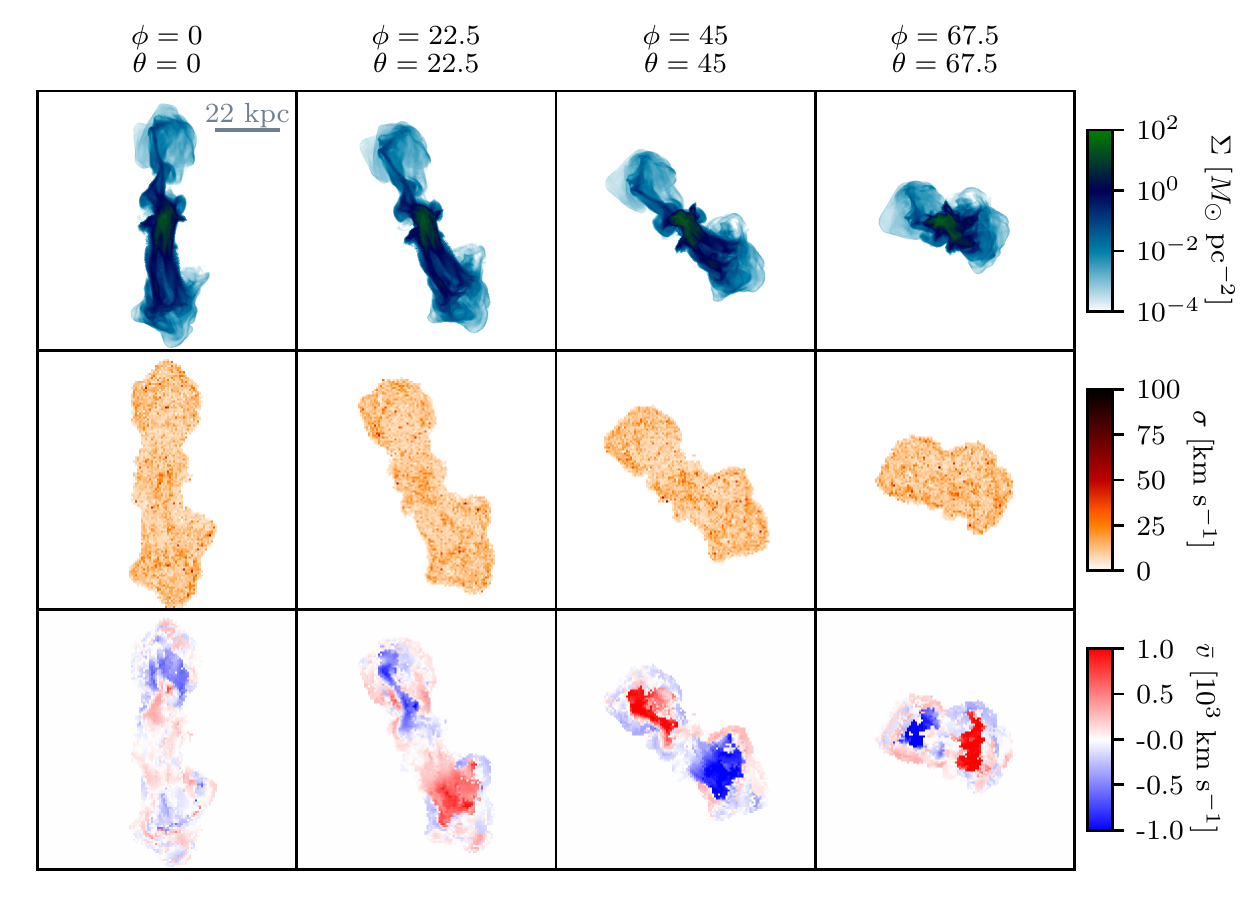}
\caption{From top to bottom, we show surface density $\Sigma$ of tracer material lifted from the center, tracer mass-weighted velocity dispersion $\sigma$ and mean velocity $\bar{v}$ along the line of sight through the simulation box. Variables are binned in pixels of $0.7\,\mathrm{kpc}\times0.7\,\mathrm{kpc}$. Images have dimensions $90\,\mathrm{kpc}\times 90\,\mathrm{kpc}$. We show results for \texttt{Fiducial} at $50\,\mathrm{Myr}$. From left to right, the jet is rotated towards the observer at angle $\theta$ and counter-clockwise in the plane at angle $\phi$. The wake accelerates the tracers up to $1000\,\mathrm{km}\,\mathrm{s}^{-1}$ along the jet axis (high $\theta$). The main velocity component shows low dispersion. Our high resolution allows us good sampling of the velocity distribution.}
    \label{fig:filaments_overview_tilt}
\end{figure*}

Our analysis is focused on the \texttt{Fiducial} run at $50\,\mathrm{Myr}$. In Figure \ref{fig:filaments_overview_tilt}, we portray the column density, velocity dispersion and mean velocity component along the line-of-sight of the traced gas in each row, respectively. The jet is rotated around the SMBH towards the observer ($\theta$) and anti-clockwise in the plane ($\phi$) at the same angle of $\theta=\phi=0^\circ,22^\circ,45^\circ,68^\circ$ in panels from left to right, respectively. At  $50\,\mathrm{Myr}$, lobes are inflated, which rise buoyantly as bubbles in the cluster. Their powerful wake causes magnetic field amplification and drags up gas from the center \citep{Jones2005,ONeill2009}. The later process is visible as filamentous structures of enhanced column densities that extend from the centre of the cluster to the bubbles.

Induced mean velocities are highest in the filamentous structures and exceed values of $1000\,\mathrm{km}\,\mathrm{s}^{-1}$. They are best visible when looking into the jet (high values of $\theta$). Furthermore, the transverse component reaches velocities exceeding $\gtrsim500\,\mathrm{km}\,\mathrm{s}^{-1}$. The inner material is dragged along with the jet while downwards motions towards the SMBH are more common in the outer parts of the cocoon. The dragged up material even penetrates the center of the bubble and the bubble morphs into a torus. High vorticity and turbulence is generated throughout the bubble and wake. However, the velocity dispersion of the dragged up material remains surprisingly low with velocities in the range $10-40\,\mathrm{km}\,\mathrm{s}^{-1}$.

\begin{figure*}
\centering

\includegraphics[width=0.48\textwidth,trim={0.2cm 0.2cm 0.2cm 0.2cm},clip]{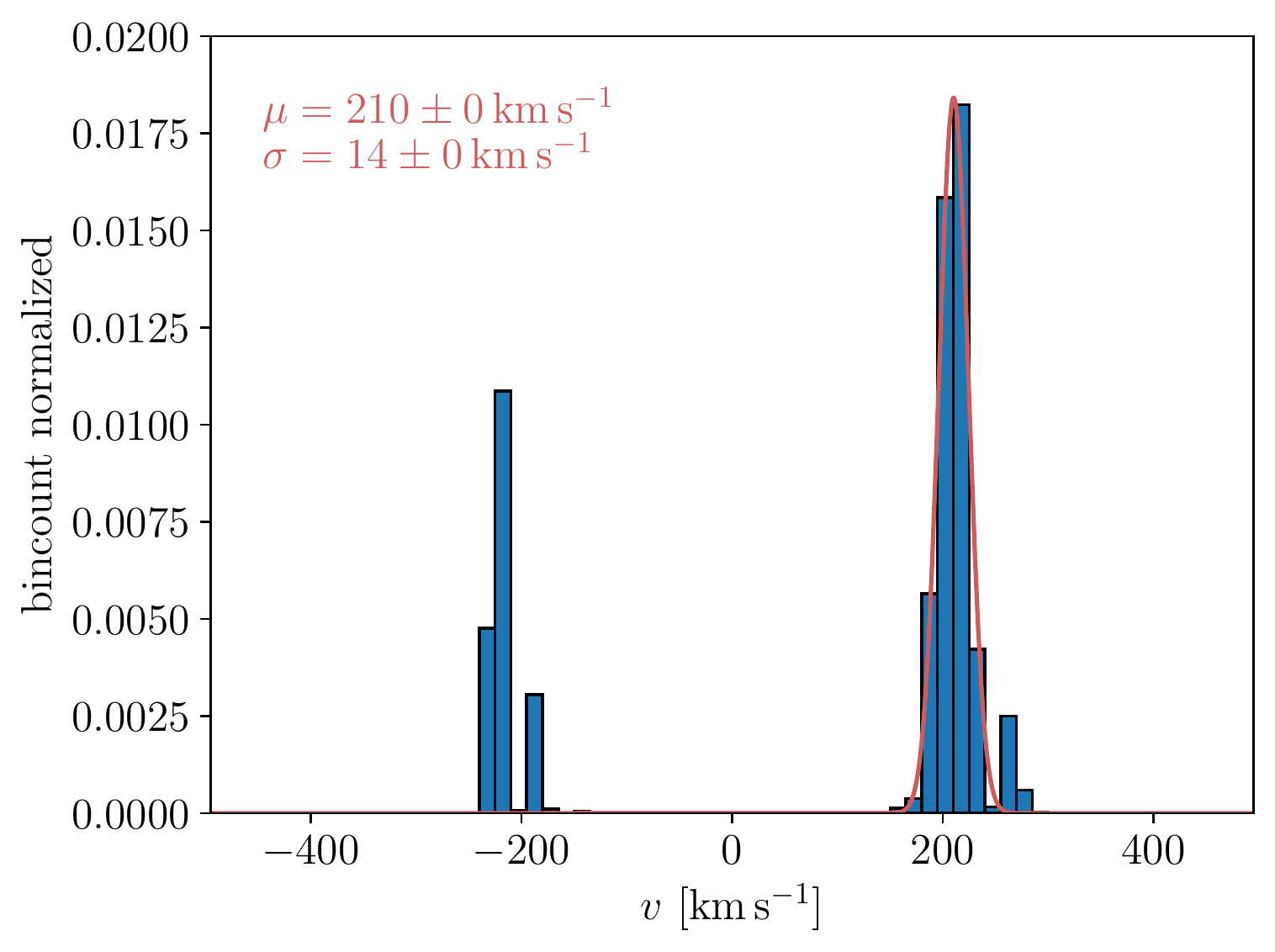}		
\includegraphics[width=0.48\textwidth,trim={0.2cm 0.2cm 0.2cm 0.2cm},clip]{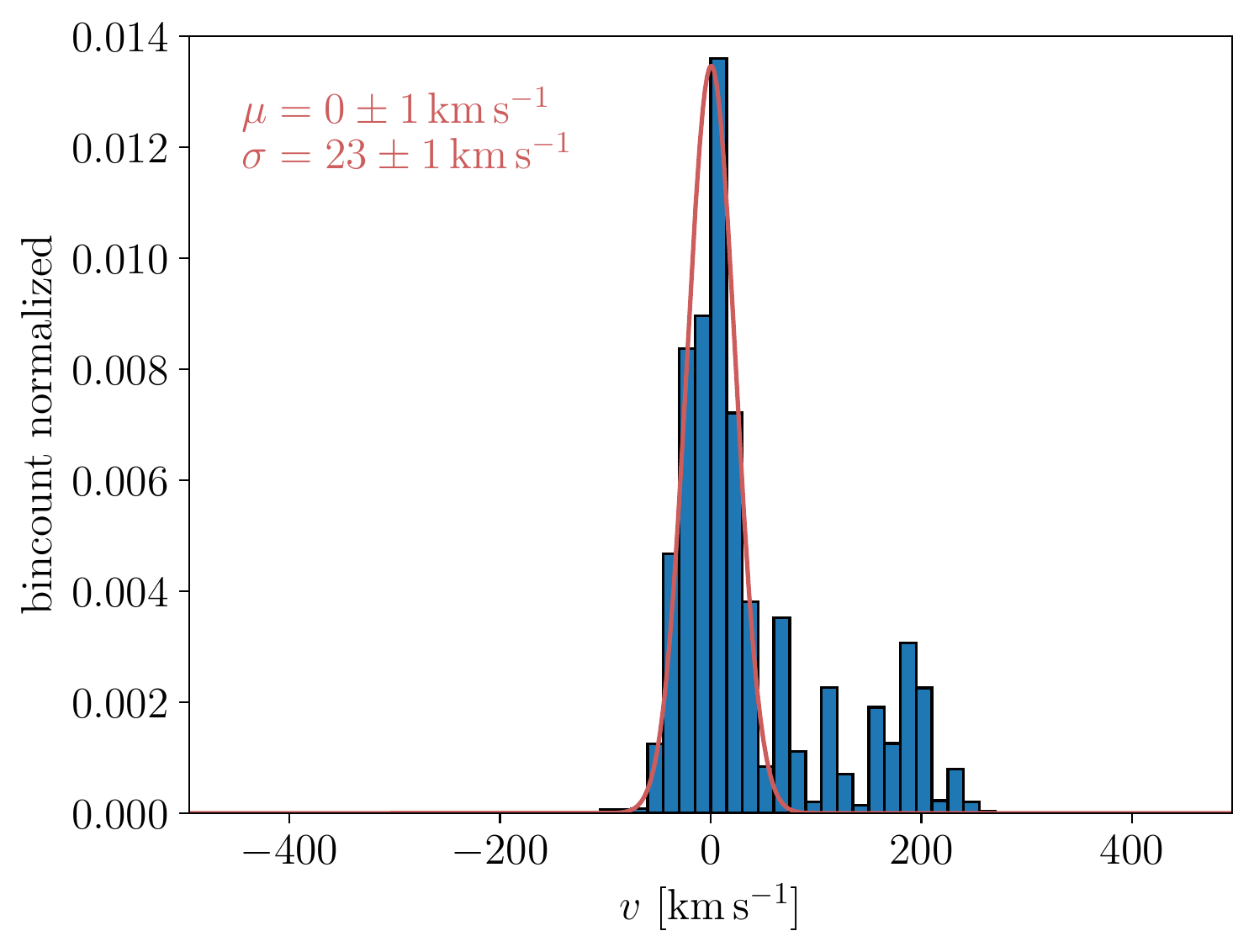}		

\includegraphics[width=0.48\textwidth,trim={0.2cm 0.2cm 0.2cm 0.2cm},clip]{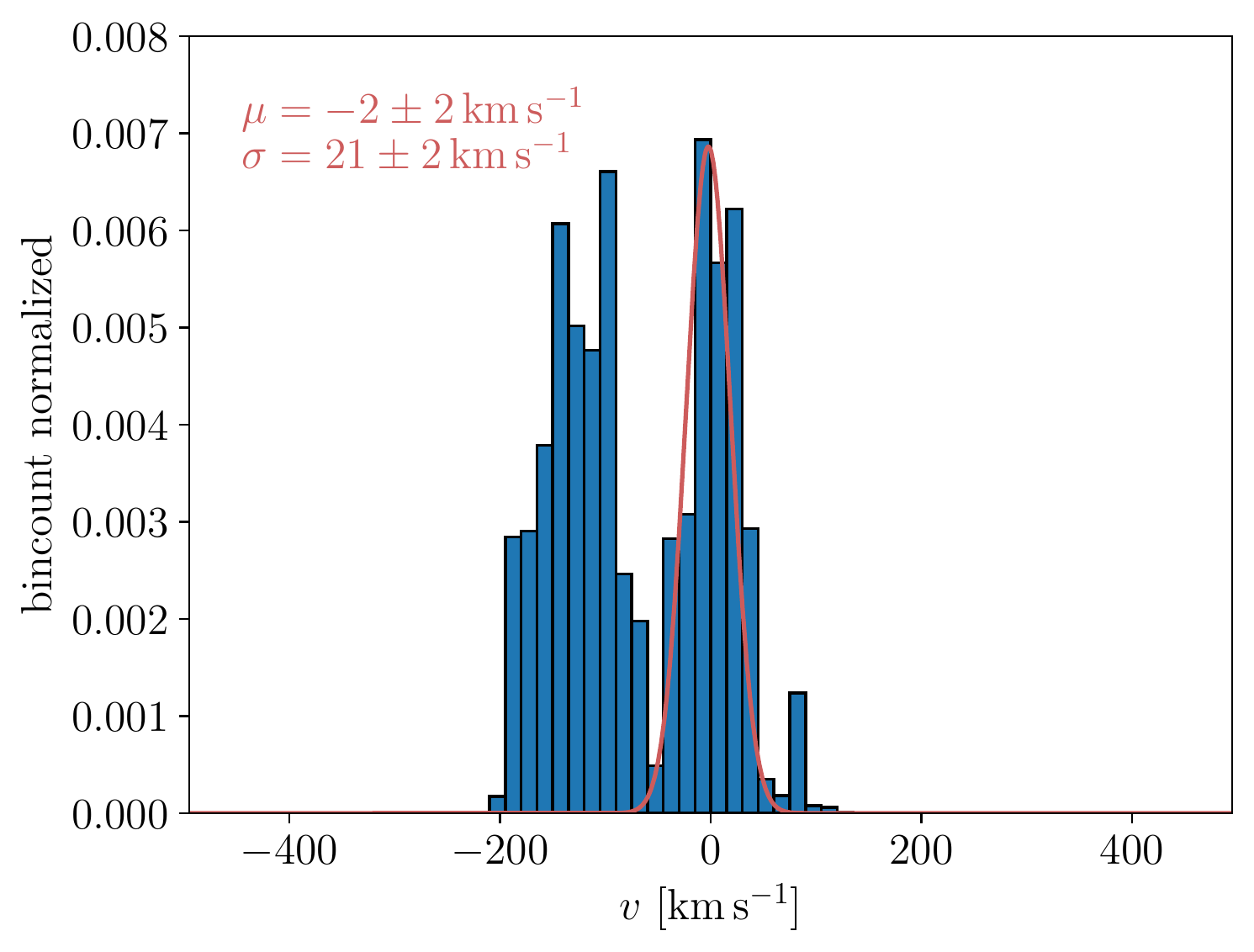}	
\includegraphics[width=0.48\textwidth,trim={0.2cm 0.2cm 0.2cm 0.2cm},clip]{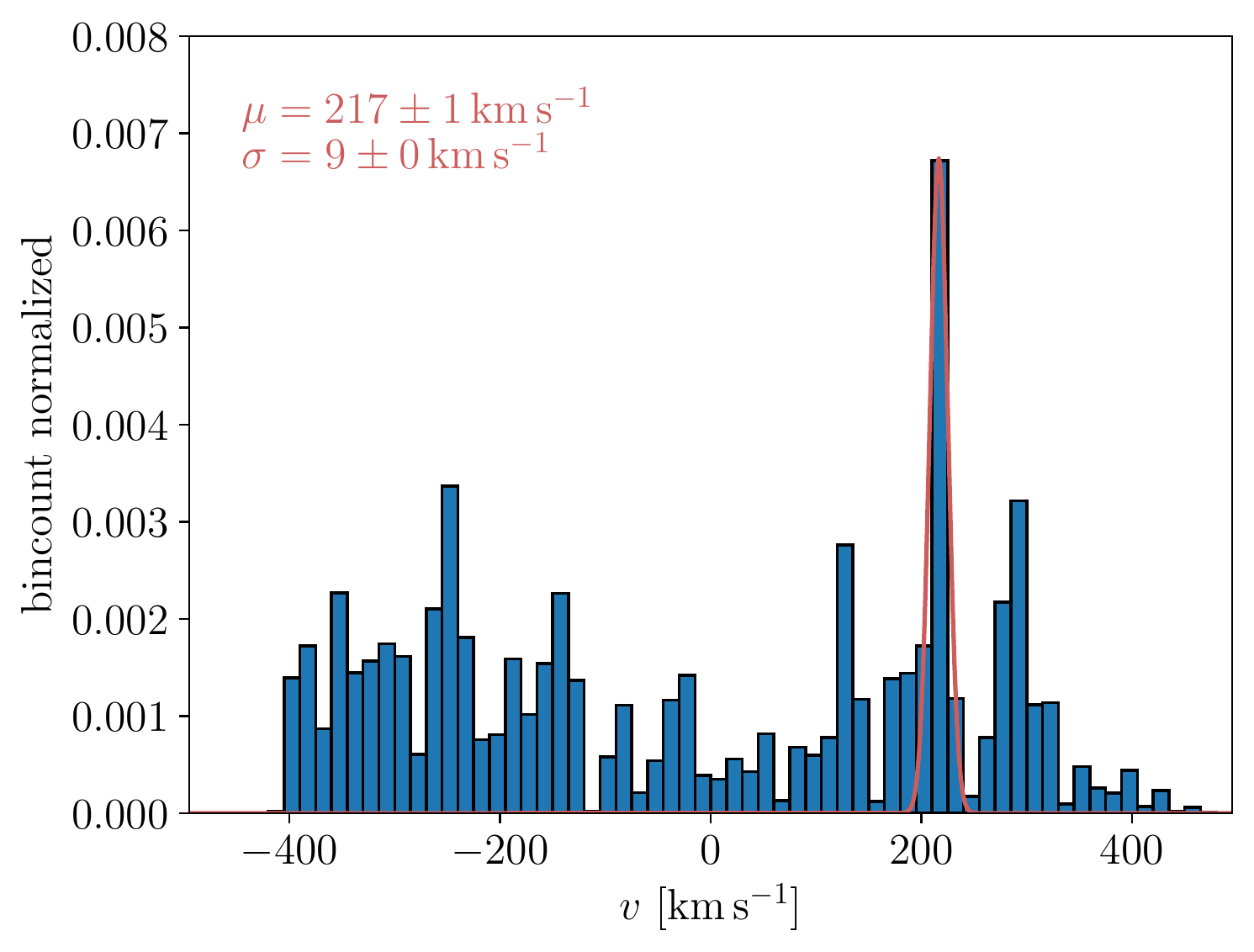}		 

\caption{Histograms of the velocity of uplifted tracers by the AGN. Panels correspond to the line-of-sight velocity distribution in a single pixel weighted with tracer mass. We also overplot Gaussian fits that are used to determine mean velocity and velocity dispersion of the main velocity component (see Figure \ref{fig:filaments_overview_tilt}). The velocity distribution is dominated by a single Gaussian in the upper panels. However, the velocity distribution in the lower panels shows two distinct peaks or a predominantly flat distribution, respectively. The induced turbulent velocities show diverse and very complex distributions. Assuming a single velocity component is clearly insufficient to represent the complex flow structure.}
    \label{fig:filaments_histograms}
\end{figure*}

Figure \ref{fig:filaments_histograms} shows a sample of four histograms that provide an overview of diversity seen in the velocity structure in individual pixels, corresponding to individual line-of-sight projections. Pixels show single Gaussian peaks with subdominant secondary flows (upper panels). At least two Gaussian components can be identified in the lower left panel. But we also find very heterogeneous velocity distributions in a significant fraction of pixels (lower right panel). Here, we see a clear peak in addition to many additional components moving in opposite directions. Thereby, a single Gaussian fit and its velocity dispersion cannot accurately account for the intermittent velocity structures of multiple velocity components in the flow patterns of the dragged-up material. Note that the turbulent ICM also influences the motion of the scalar. However, our comparison of induced velocities in \texttt{Fiducial} and \texttt{NoTurb} showed that the ICM contribution is secondary. 

In summary, we are left with a remarkably coherent outflow in the wake of the bubble, at or exceeding the buoyant rising velocity of the jet-inflated bubble, reaching Mach numbers of almost unity (see Figure~\ref{fig:velocity_impact_jet}, bottom panel, sound speed around $10^3\,\mathrm{km}\,\mathrm{s}^{-1}$), and fairly insensitive to preexisting turbulence. Since the bubble velocity is mostly set by cluster properties and bubble size \citep{Churazov2001}, we expect it to be fairly insensitive to details of the jet other than the total injected energy \citep[which determines the bubble size, as shown in][]{Ehlert2018}. In the outflow reference frame, the local velocity dispersions constitute Mach numbers of only around 0.01-0.05. This has important implications for future studies of the thermodynamics of these outflows since it allows to study thermal instability in local simulations of outward moving patches of gas, without the need for a global, cluster wide simulation.  While we only investigate one specific case, there is no reason to assume this changes qualitatively with changed jet parameters, though it is plausible to assume that the range of turbulent Mach numbers in the outflow frame across varying jet parameters is larger than presented here.

\section{Conclusions}
\label{sec:conclusion}
We run MHD simulations of jet-inflated bubbles in a Perseus-like CC cluster with a turbulent magnetic field. There is an inherent relation between large-scale magnetic and velocity fields, which we connect to two observables:
line of sight velocity dispersion as measured by high spectral resolution X-ray spectroscopy and RM. Furthermore, we study the influence of the AGN on the velocity field, its detectability and the effect of spatial resolution. Finally, we analyze the velocity structure of uplifted central gas in the wake of the buoyantly rising radio-lobes. We summarize our findings as follows.

We find Faraday RMs that are an order of magnitude above observed values. This is somewhat surprising as magnetic field parameters were directly taken from observations. The magnetic field of the jet has negligible effect on the overall RM (see Section \ref{sec:varymagneticfieldstrength}). We identify four effects that could be responsible for the discrepancy. (i) Possibly our adopted coherence scale is too large and we therefore underpredict depolarization. To test this hypothesis, we ran an additional simulation with an injetion scale of $k_\mathrm{inj}=15^{-1}\,\mathrm{kpc}^{-1}$, which decreases RM by $30\,\%$ (see Figure \ref{fig:rotation_measure_injection_scale}). (ii) Observed RMs are limited to few small patches in clusters that are provided by the angular extends of radio lobes. Especially the bright central radio sources in CC clusters imply a large dynamic range that challenges high-frequency polarized observations of lobes, which are negligibly affected by Faraday depolarisation. (iii) Beam smoothing artificially lowers the dispersion of observed RMs. We neglect this effect here. (iv) \cite{Vazza2018a} find their simulated magnetic fields to depart from a Gaussian distribution that is usually assumed when modeling Faraday RMs. Consequently, observations of RM possibly overestimate cluster magnetic field strengths. This highlights the relevance of cosmological MHD simulations that are able to self-consistently drive and sustain large scale magnetic fields. Moreover, this calls for a dedicated synthetic modeling of observations to take into account all possible observational effects.

\begin{itemize}
\item Measurements of large scale magnetic fields and velocities in galaxy clusters are inherently coupled. The strength, radial scaling and injection scale of our magnetic fields are motivated by observations. Interestingly, they yield RM values that are an order of magnitude higher than observed. However, observationally obtained magnetic field strengths may suffer from Faraday depolarization and only probe a very limited spatial window. Assuming too large magnetic coherence scales may also bias the inferred magnetic field strengths high. Finally, beam smoothing and possible departures from Gaussianity may further alleviate this discrepancy. Future work is needed to address this open problem.
\item The influence of the AGN driven jet on the velocity fields is limited to the lobe's proximity. Gas flows on scales larger than the characteristic size of the lobes are not AGN driven.
\item Given the arcmin resoution of the analysis of the \textit{Hitomi} data of Perseus, we conclude that the measured  cluster turbulence is likely not directly driven by rising radio lobes.
\item Central gas is dragged up in filament-like structures by the AGN. A complex multi-component velocity structure is induced in the lifted material. While the bulk velocity of the lifted material exceeds $1000$ km~s$^{-1}$, the line of sight velocity dispersion is remarkably low with only about $50$ km~s$^{-1}$. The previously central gas remains constrained to the wake and near proximity of the rising bubbles.
\end{itemize}

The connection of line of sight velocity dispersion and RM studied 
in this work highlights the multi-facetted nature of intra-cluster turbulence, and 
the role numerical simulations can play to connect the different observables.
We restricted this study to non-radiative simulations in order to quantify the
role of 'stirring' by rising radio lobes without the additional effects of an 
(AGN moderated) cooling flow, as well as to an individual outburst. In the future, we plan to study the influence of cooling on the local AGN-induced velocity fields. This will allow us to compare our model to observed filaments in CC clusters.

\section*{Acknowledgements}
KE and RW would like to thank John ZuHone for the insightful discussions and for making the \textsc{pyxsim} code publicly available. KE and CP acknowledge support by the European Research Council under ERC-CoG grant CRAGSMAN-646955.

\section*{Data availability}
The data underlying this article will be shared upon request to the
corresponding author.

%%%%%%%%%%%%%%%%%%%%%%%%%%%%%%%%%%%%%%%%%%%%%%%%%%

%%%%%%%%%%%%%%%%%%%% REFERENCES %%%%%%%%%%%%%%%%%%

\bibliographystyle{mnras}

\bibliography{library}

\begin{thebibliography}{}
\makeatletter
\relax
\def\mn@urlcharsother{\let\do\@makeother \do\$\do\&\do\#\do\^\do\_\do\%\do\~}
\def\mn@doi{\begingroup\mn@urlcharsother \@ifnextchar [ {\mn@doi@}
  {\mn@doi@[]}}
\def\mn@doi@[#1]#2{\def\@tempa{#1}\ifx\@tempa\@empty \href
  {http://dx.doi.org/#2} {doi:#2}\else \href {http://dx.doi.org/#2} {#1}\fi
  \endgroup}
\def\mn@eprint#1#2{\mn@eprint@#1:#2::\@nil}
\def\mn@eprint@arXiv#1{\href {http://arxiv.org/abs/#1} {{\tt arXiv:#1}}}
\def\mn@eprint@dblp#1{\href {http://dblp.uni-trier.de/rec/bibtex/#1.xml}
  {dblp:#1}}
\def\mn@eprint@#1:#2:#3:#4\@nil{\def\@tempa {#1}\def\@tempb {#2}\def\@tempc
  {#3}\ifx \@tempc \@empty \let \@tempc \@tempb \let \@tempb \@tempa \fi \ifx
  \@tempb \@empty \def\@tempb {arXiv}\fi \@ifundefined
  {mn@eprint@\@tempb}{\@tempb:\@tempc}{\expandafter \expandafter \csname
  mn@eprint@\@tempb\endcsname \expandafter{\@tempc}}}

\bibitem[\protect\citeauthoryear{Asplund, Grevesse, Sauval  \& Scott}{Asplund
  et~al.}{2009}]{Asplund2009}
Asplund M.,  Grevesse N.,  Sauval A.~J.,   Scott P.,  2009, \mn@doi [Annu. Rev.
  Astron. Astrophys.] {10.1146/annurev.astro.46.060407.145222}, 47, 481

\bibitem[\protect\citeauthoryear{Bambic \& Reynolds}{Bambic \&
  Reynolds}{2019}]{Bambic2019}
Bambic C.~J.,  Reynolds C.~S.,  2019, \mn@doi [Astrophys. J.]
  {10.3847/1538-4357/ab4daf}, 886, 78

\bibitem[\protect\citeauthoryear{Bambic, Morsony  \& Reynolds}{Bambic
  et~al.}{2018}]{Bambic2018}
Bambic C.~J.,  Morsony B.~J.,   Reynolds C.~S.,  2018, Astrophys. J., 857, 84

\bibitem[\protect\citeauthoryear{Beckmann et~al.,}{Beckmann
  et~al.}{2019}]{Beckmann2019a}
Beckmann R.~S.,  et~al., 2019, \mn@doi [Astron. Astrophys.]
  {10.1051/0004-6361/201936188}, 631, 1

\bibitem[\protect\citeauthoryear{Beresnyak}{Beresnyak}{2012}]{Beresnyak2012}
Beresnyak A.,  2012, \mn@doi [Phys. Scr.] {10.1088/0031-8949/86/05/058201}, 86,
  058201

\bibitem[\protect\citeauthoryear{Beresnyak \& Miniati}{Beresnyak \&
  Miniati}{2016}]{Beresnyak2016}
Beresnyak A.,  Miniati F.,  2016, \mn@doi [Astrophys. J.]
  {10.3847/0004-637x/817/2/127}, 817, 127

\bibitem[\protect\citeauthoryear{B{\^{i}}rzan, Rafferty, McNamara, Wise  \&
  Nulsen}{B{\^{i}}rzan et~al.}{2004}]{Birzan2004}
B{\^{i}}rzan L.,  Rafferty D.~A.,  McNamara B.~R.,  Wise M.~W.,   Nulsen P.
  E.~J.,  2004, \mn@doi [Astrophys. J.] {10.1086/383519}, 607, 800

\bibitem[\protect\citeauthoryear{Bonafede, Feretti, Murgia, Govoni, Giovannini,
  Dallacasa, Dolag  \& Taylor}{Bonafede et~al.}{2010}]{Bonafede2010}
Bonafede A.,  Feretti L.,  Murgia M.,  Govoni F.,  Giovannini G.,  Dallacasa
  D.,  Dolag K.,   Taylor G.~B.,  2010, \mn@doi [Astron. Astrophys.]
  {10.1051/0004-6361/200913696}, 513, A30

\bibitem[\protect\citeauthoryear{Bourne \& Sijacki}{Bourne \&
  Sijacki}{2017}]{Bourne2017}
Bourne M.~A.,  Sijacki D.,  2017, \mn@doi [Mon. Not. R. Astron. Soc.]
  {10.1093/mnras/stx2269}, 472, 4707

\bibitem[\protect\citeauthoryear{Bourne \& Sijacki}{Bourne \&
  Sijacki}{2020}]{Bourne2020}
Bourne A.,  Sijacki D.,  2020, {AGN jet feedback on a moving mesh: weak shocks
  and lobe disruption gently prevent the cooling catastrophe}.
preprint (arXiv:2008.12784) (\mn@eprint {arXiv} {arXiv:2008.12784v1})

\bibitem[\protect\citeauthoryear{Bourne, Sijacki  \& Puchwein}{Bourne
  et~al.}{2019}]{Bourne2019}
Bourne M.~A.,  Sijacki D.,   Puchwein E.,  2019, \mn@doi [Mon. Not. R. Astron.
  Soc.] {10.1093/mnras/stz2604}, 490, 343

\bibitem[\protect\citeauthoryear{Bulbul, Smith, Foster, Cottam, Loewenstein,
  Mushotzky  \& Shafer}{Bulbul et~al.}{2012}]{Bulbul2012}
Bulbul G.~E.,  Smith R.~K.,  Foster A.,  Cottam J.,  Loewenstein M.,  Mushotzky
  R.,   Shafer R.,  2012, \mn@doi [Astrophys. J.] {10.1088/0004-637X/747/1/32},
  747, 32

\bibitem[\protect\citeauthoryear{Chen, Heinz  \& En{\ss}lin}{Chen
  et~al.}{2019}]{Chen2019}
Chen Y.-H.~H.,  Heinz S.,   En{\ss}lin T.~A.,  2019, \mn@doi [Mon. Not. R.
  Astron. Soc.] {10.1093/mnras/stz2256}, 11, 1939

\bibitem[\protect\citeauthoryear{Cho}{Cho}{2014}]{Cho2014}
Cho J.,  2014, \mn@doi [Astrophys. J.] {10.1088/0004-637X/797/2/133}, 797

\bibitem[\protect\citeauthoryear{Churazov, Br{\"{u}}ggen, Kaiser,
  B{\"{o}}hringer  \& Forman}{Churazov et~al.}{2001}]{Churazov2001}
Churazov E.,  Br{\"{u}}ggen M.,  Kaiser C.~R.,  B{\"{o}}hringer H.,   Forman
  W.,  2001, \mn@doi [Astrophys. J.] {10.1086/321357}, 554, 261

\bibitem[\protect\citeauthoryear{Churazov, Forman, Jones  \&
  B{\"{o}}hringer}{Churazov et~al.}{2003}]{Churazov2003}
Churazov E.,  Forman W.,  Jones C.,   B{\"{o}}hringer H.,  2003, \mn@doi
  [Astrophys. J.] {10.1086/374923}, 590, 225

\bibitem[\protect\citeauthoryear{Churazov et~al.,}{Churazov
  et~al.}{2012}]{Churazov2012}
Churazov E.,  et~al., 2012, \mn@doi [Mon. Not. R. Astron. Soc.]
  {10.1111/j.1365-2966.2011.20372.x}, 421, 1123

\bibitem[\protect\citeauthoryear{Clarke}{Clarke}{2004}]{Clarke2004a}
Clarke T.~E.,  2004, \mn@doi [To Appear a Dedic. issue J. Korean Astron. Soc.
  (JKAS). Proc. "International Conf. Cosm. Rays Magn. Fields Large Scale
  Struct.] {10.5303/JKAS.2004.37.5.337}, p.~6

\bibitem[\protect\citeauthoryear{{De Plaa}, Zhuravleva, Werner, Kaastra,
  Churazov, Smith, Raassen  \& Grange}{{De Plaa} et~al.}{2012}]{DePlaa2012}
{De Plaa} J.,  Zhuravleva I.,  Werner N.,  Kaastra J.~S.,  Churazov E.,  Smith
  R.~K.,  Raassen A.~J.,   Grange Y.~G.,  2012, \mn@doi [Astron. Astrophys.]
  {10.1051/0004-6361/201118404}, 539, A34

\bibitem[\protect\citeauthoryear{Diehl, Li, Fryer  \& Rafferty}{Diehl
  et~al.}{2008}]{Diehl2008}
Diehl S.,  Li H.,  Fryer C.~L.,   Rafferty D.,  2008, \mn@doi [Astrophys. J.]
  {10.1086/591310}, 687, 173

\bibitem[\protect\citeauthoryear{Dolag, Vazza, Brunetti  \& Tormen}{Dolag
  et~al.}{2005}]{Dolag2005}
Dolag K.,  Vazza F.,  Brunetti G.,   Tormen G.,  2005, \mn@doi [Mon. Not. R.
  Astron. Soc.] {10.1111/j.1365-2966.2005.09630.x}, 364, 753

\bibitem[\protect\citeauthoryear{Dominguez-Fernandez, Vazza, Br{\"{u}}ggen  \&
  Brunetti}{Dominguez-Fernandez et~al.}{2019}]{Dominguez-Fernandez2019}
Dominguez-Fernandez P.,  Vazza F.,  Br{\"{u}}ggen M.,   Brunetti G.,  2019,
  \mn@doi [Mon. Not. R. Astron. Soc.] {10.1093/mnras/stz877}, 486, 623

\bibitem[\protect\citeauthoryear{Donnert, Dolag, Lesch  \&
  M{\"{u}}ller}{Donnert et~al.}{2009}]{Donnert2009}
Donnert J.,  Dolag K.,  Lesch H.,   M{\"{u}}ller E.,  2009, \mn@doi [Mon. Not.
  R. Astron. Soc.] {10.1111/j.1365-2966.2008.14132.x}, 392, 1008

\bibitem[\protect\citeauthoryear{Donnert, Vazza, Br{\"{u}}ggen  \&
  ZuHone}{Donnert et~al.}{2018}]{Donnert2018}
Donnert J.,  Vazza F.,  Br{\"{u}}ggen M.,   ZuHone J.,  2018, \mn@doi [Space
  Sci. Rev.] {10.1007/s11214-018-0556-8}, 214, 122

\bibitem[\protect\citeauthoryear{Dubois, Devriendt, Slyz  \& Silk}{Dubois
  et~al.}{2009}]{Dubois2009}
Dubois Y.,  Devriendt J.,  Slyz A.,   Silk J.,  2009, \mn@doi [Mon. Not. R.
  Astron. Soc. Lett.] {10.1111/j.1745-3933.2009.00721.x}, 399, 49

\bibitem[\protect\citeauthoryear{Ehlert, Weinberger, Pfrommer, Pakmor  \&
  Springel}{Ehlert et~al.}{2018}]{Ehlert2018}
Ehlert K.,  Weinberger R.,  Pfrommer C.,  Pakmor R.,   Springel V.,  2018,
  \mn@doi [Mon. Not. R. Astron. Soc.] {10.1093/mnras/sty2397}, 481, 2878

\bibitem[\protect\citeauthoryear{En{\ss}lin, Pfrommer, Miniati  \&
  Subramanian}{En{\ss}lin et~al.}{2011}]{Ensslin2011}
En{\ss}lin T.~A.,  Pfrommer C.,  Miniati F.,   Subramanian K.,  2011, \mn@doi
  [Astron. Astrophys.] {10.1051/0004-6361/201015652}, 527, A99

\bibitem[\protect\citeauthoryear{Fabian}{Fabian}{2012}]{Fabian2012}
Fabian A.~C.,  2012, \mn@doi [Annu. Rev. Astron. Astrophys.]
  {10.1146/annurev-astro-081811-125521}, 50, 455

\bibitem[\protect\citeauthoryear{Fabian, Walker, Russell, Pinto, Sanders  \&
  Reynolds}{Fabian et~al.}{2017}]{Fabian2017}
Fabian A.~C.,  Walker S.~A.,  Russell H.~R.,  Pinto C.,  Sanders J.~S.,
  Reynolds C.~S.,  2017, \mn@doi [Mon. Not. R. Astron. Soc.]
  {10.1093/mnrasl/slw170}, 464, L1

\bibitem[\protect\citeauthoryear{Fujita, Cen  \& Zhuravleva}{Fujita
  et~al.}{2020}]{Fujita2019}
Fujita Y.,  Cen R.,   Zhuravleva I.,  2020, \mn@doi [Mon. Not. R. Astron. Soc.]
  {10.1093/mnras/staa1087}, 494, 5507

\bibitem[\protect\citeauthoryear{Gaspari \& Churazov}{Gaspari \&
  Churazov}{2013}]{Gaspari2013a}
Gaspari M.,  Churazov E.,  2013, \mn@doi [Astron. Astrophys.]
  {10.1051/0004-6361/201322295}, 559

\bibitem[\protect\citeauthoryear{Gaspari, Ruszkowski  \& Sharma}{Gaspari
  et~al.}{2012}]{Gaspari2012}
Gaspari M.,  Ruszkowski M.,   Sharma P.,  2012, \mn@doi [Astrophys. J.]
  {10.1088/0004-637X/746/1/94}, 746, 94

\bibitem[\protect\citeauthoryear{Gendron-Marsolais et~al.,}{Gendron-Marsolais
  et~al.}{2018}]{Gendron-Marsolais2018}
Gendron-Marsolais M.,  et~al., 2018, Mon. Not. R. Astron. Soc., 479, L28

\bibitem[\protect\citeauthoryear{Gitti, Brighenti  \& McNamara}{Gitti
  et~al.}{2012}]{Gitti2012}
Gitti M.,  Brighenti F.,   McNamara B.~R.,  2012, \mn@doi [Adv. Astron.]
  {10.1155/2012/950641}, 2012

\bibitem[\protect\citeauthoryear{Govoni et~al.,}{Govoni
  et~al.}{2010}]{Govoni2010}
Govoni F.,  et~al., 2010, \mn@doi [Astron. Astrophys.]
  {10.1051/0004-6361/200913665}, 522, 1

\bibitem[\protect\citeauthoryear{Govoni et~al.,}{Govoni
  et~al.}{2017}]{Govoni2017}
Govoni F.,  et~al., 2017, \mn@doi [Astron. Astrophys.]
  {10.1051/0004-6361/201630349}, 603, A122

\bibitem[\protect\citeauthoryear{Gronke \& Oh}{Gronke \& Oh}{2018}]{Gronke2018}
Gronke M.,  Oh S.~P.,  2018, \mn@doi [Mon. Not. R. Astron. Soc.]
  {10.1093/mnrasl/sly131}, 480, L111

\bibitem[\protect\citeauthoryear{Gu, Zhuravleva, Churazov, Paerels, Kaastra  \&
  Yamaguchi}{Gu et~al.}{2018}]{Gu2018}
Gu L.,  Zhuravleva I.,  Churazov E.,  Paerels F.,  Kaastra J.,   Yamaguchi H.,
  2018, \mn@doi [Space Sci. Rev.] {10.1007/s11214-018-0544-z}, 214

\bibitem[\protect\citeauthoryear{Guidetti, Murgia, Govoni, Parma, Gregorini, de
  Ruiter, Cameron  \& Fanti}{Guidetti et~al.}{2008}]{Guidetti2008}
Guidetti D.,  Murgia M.,  Govoni F.,  Parma P.,  Gregorini L.,  de Ruiter H.,
  Cameron R.~A.,   Fanti R.,  2008, \mn@doi [Astron. Astrophys.]
  {10.1051/0004-6361:20065964}, 483, 699

\bibitem[\protect\citeauthoryear{Guidetti, Laing, Croston, Bridle  \&
  Parma}{Guidetti et~al.}{2012}]{Guidetti2012}
Guidetti D.,  Laing R.~A.,  Croston J.~H.,  Bridle A.~H.,   Parma P.,  2012,
  \mn@doi [Mon. Not. R. Astron. Soc.] {10.1111/j.1365-2966.2012.20961.x}, 423,
  1335

\bibitem[\protect\citeauthoryear{Guo \& Oh}{Guo \& Oh}{2008}]{Guo2008}
Guo F.,  Oh S.~P.,  2008, \mn@doi [Mon. Not. R. Astron. Soc.]
  {10.1111/j.1365-2966.2007.12692.x}, 384, 251

\bibitem[\protect\citeauthoryear{Guo, Duan  \& Yuan}{Guo
  et~al.}{2018}]{Guo2018}
Guo F.,  Duan X.,   Yuan Y.~F.,  2018, \mn@doi [Mon. Not. R. Astron. Soc.]
  {10.1093/MNRAS/STX2404}, 473, 1332

\bibitem[\protect\citeauthoryear{Han}{Han}{2017}]{Han2017}
Han J.,  2017, \mn@doi [Annu. Rev. Astron. Astrophys.]
  {10.1146/annurev-astro-091916-055221}, 55, 111

\bibitem[\protect\citeauthoryear{Hillel \& Soker}{Hillel \&
  Soker}{2017a}]{Hillel2017a}
Hillel S.,  Soker N.,  2017a, \mn@doi [Mon. Not. R. Astron. Soc. Lett.]
  {10.1093/mnrasl/slw231}, 466, L39

\bibitem[\protect\citeauthoryear{Hillel \& Soker}{Hillel \&
  Soker}{2017b}]{Hillel2017}
Hillel S.,  Soker N.,  2017b, \mn@doi [Astrophys. J.]
  {10.3847/1538-4357/aa81c5}, 845, 91

\bibitem[\protect\citeauthoryear{Hillel \& Soker}{Hillel \&
  Soker}{2018}]{Hillel2018}
Hillel S.,  Soker N.,  2018, \mn@doi [Res. Astron. Astrophys.] {10.1088/1674},
  18, 81

\bibitem[\protect\citeauthoryear{{Hitomi Collaboration}}{{Hitomi
  Collaboration}}{2016}]{HitomiCollaboration2016}
{Hitomi Collaboration} 2016, \mn@doi [Nature] {10.1038/nature18627}, 535, 117

\bibitem[\protect\citeauthoryear{{Hitomi Collaboration}}{{Hitomi
  Collaboration}}{2018}]{HitomiCollaboration2018}
{Hitomi Collaboration} 2018, \mn@doi [Publ. Astron. Soc. Japan]
  {10.1093/pasj/psx138}, 70, 9

\bibitem[\protect\citeauthoryear{Jacob \& Pfrommer}{Jacob \&
  Pfrommer}{2017a}]{Jacob2016a}
Jacob S.,  Pfrommer C.,  2017a, \mn@doi [Mon. Not. R. Astron. Soc.]
  {10.1093/mnras/stx132}, 467, 1449

\bibitem[\protect\citeauthoryear{Jacob \& Pfrommer}{Jacob \&
  Pfrommer}{2017b}]{Jacob2016b}
Jacob S.,  Pfrommer C.,  2017b, \mn@doi [Mon. Not. R. Astron. Soc.]
  {10.1093/mnras/stx132}, 467, 1478

\bibitem[\protect\citeauthoryear{Jones \& {De Young}}{Jones \& {De
  Young}}{2005}]{Jones2005}
Jones T.~W.,  {De Young} D.~S.,  2005, \mn@doi [Astrophys. J.]
  {10.1086/429157}, 624, 586

\bibitem[\protect\citeauthoryear{Kazantsev}{Kazantsev}{1968}]{Kazantsev1968}
Kazantsev A.,  1968, Sov. J. Exp. Theor. Phys., 26, 1031

\bibitem[\protect\citeauthoryear{Koekemoer, O'Dea, Sarazin, McNamara, Donahue,
  Voit, Baum  \& Gallimore}{Koekemoer et~al.}{1999}]{Koekemoer1999}
Koekemoer A.~M.,  O'Dea C.~P.,  Sarazin C.~L.,  McNamara B.~R.,  Donahue M.,
  Voit G.~M.,  Baum S.~A.,   Gallimore J.~F.,  1999, ApJ, 525, 621

\bibitem[\protect\citeauthoryear{Kuchar \& En{\ss}lin}{Kuchar \&
  En{\ss}lin}{2011}]{Kuchar2011}
Kuchar P.,  En{\ss}lin T.~A.,  2011, \mn@doi [Astron. Astrophys.]
  {10.1051/0004-6361/200913918}, 529, 13

\bibitem[\protect\citeauthoryear{Kulsrud}{Kulsrud}{2005}]{Kulsrud2005}
Kulsrud R.~M.,  2005, {Plasma Physics for Astrophysics}.
Princeton University Press, Princeton, NJ

\bibitem[\protect\citeauthoryear{Lau, Gaspari, Nagai  \& Coppi}{Lau
  et~al.}{2017}]{Lau2017}
Lau E.~T.,  Gaspari M.,  Nagai D.,   Coppi P.,  2017, \mn@doi [Astrophys. J.]
  {10.3847/1538-4357/aa8c00}, 849, 54

\bibitem[\protect\citeauthoryear{Li, Ruszkowski  \& Bryan}{Li
  et~al.}{2017}]{Li2017}
Li Y.,  Ruszkowski M.,   Bryan G.~L.,  2017, \mn@doi [Astrophys. J.]
  {10.3847/1538-4357/aa88c1}, 847, 106

\bibitem[\protect\citeauthoryear{Li, Hopkins, Squire  \& Hummels}{Li
  et~al.}{2020}]{Li_Hopkins2020}
Li Z.,  Hopkins P.~F.,  Squire J.,   Hummels C.,  2020, \mn@doi [Mon. Not. R.
  Astron. Soc.] {10.1093/mnras/stz3567}, 492, 1841

\bibitem[\protect\citeauthoryear{Loewenstein, Zweibel  \& Begelman}{Loewenstein
  et~al.}{1991}]{Loewenstein1991}
Loewenstein M.,  Zweibel E.~G.,   Begelman M.~C.,  1991, Astrophys. J., 377,
  392

\bibitem[\protect\citeauthoryear{Martizzi, Quataert, Faucher-Gigu{\`{e}}re  \&
  Fielding}{Martizzi et~al.}{2019}]{Martizzi2018}
Martizzi D.,  Quataert E.,  Faucher-Gigu{\`{e}}re C.~A.,   Fielding D.,  2019,
  \mn@doi [Mon. Not. R. Astron. Soc.] {10.1093/mnras/sty3273}, 483, 2465

\bibitem[\protect\citeauthoryear{Martz et~al.,}{Martz et~al.}{2020}]{Martz2020}
Martz C.~G.,  et~al., 2020, \mn@doi [Astrophys. J.] {10.3847/1538-4357/ab96cd},
  897, 57

\bibitem[\protect\citeauthoryear{McNamara \& Nulsen}{McNamara \&
  Nulsen}{2012}]{McNamara2012}
McNamara B.~R.,  Nulsen P. E.~J.,  2012, \mn@doi [New J. Phys.]
  {10.1088/1367-2630/14/5/055023}, 14, 40

\bibitem[\protect\citeauthoryear{McNamara, Russell, Nulsen, Hogan, Fabian,
  Pulido  \& Edge}{McNamara et~al.}{2016}]{McNamara2016}
McNamara B.~R.,  Russell H.~R.,  Nulsen P. E.~J.,  Hogan M.~T.,  Fabian A.~C.,
  Pulido F.,   Edge A.~C.,  2016, \mn@doi [Astrophys. J.]
  {10.3847/0004-637X/830/2/79}, 830, 1

\bibitem[\protect\citeauthoryear{Mohapatra \& Sharma}{Mohapatra \&
  Sharma}{2019}]{Mohapatra2019}
Mohapatra R.,  Sharma P.,  2019, \mn@doi [Mon. Not. R. Astron. Soc.]
  {10.1093/mnras/stz328}, 484, 4881

\bibitem[\protect\citeauthoryear{Mukherjee, Bicknell, Sutherland  \&
  Wagner}{Mukherjee et~al.}{2016}]{Mukherjee2016}
Mukherjee D.,  Bicknell G.~V.,  Sutherland R.,   Wagner A.,  2016, \mn@doi
  [Mon. Not. R. Astron. Soc.] {10.1093/mnras/stw1368}, 461, 967

\bibitem[\protect\citeauthoryear{Mukherjee, Bodo, Mignone, Rossi  \&
  Vaidya}{Mukherjee et~al.}{2020}]{Mukherjee2020}
Mukherjee D.,  Bodo G.,  Mignone A.,  Rossi P.,   Vaidya B.,  2020, \mn@doi
  [Mon. Not. R. Astron. Soc.] {10.1093/mnras/staa2934}, 499, 681

\bibitem[\protect\citeauthoryear{Navarro, Frenk  \& White}{Navarro
  et~al.}{1996}]{Navarro1996}
Navarro J.~F.,  Frenk C.~S.,   White S. D.~M.,  1996, \mn@doi [ApJ]
  {10.1086/177173}, 462, 563

\bibitem[\protect\citeauthoryear{Navarro, Frenk  \& White}{Navarro
  et~al.}{1997}]{Navarro1997}
Navarro J.~F.,  Frenk C.~S.,   White S. D.~M.,  1997, \mn@doi [ApJ]
  {10.1086/304888}, 490, 493

\bibitem[\protect\citeauthoryear{O'Neill, {De Young}  \& Jones}{O'Neill
  et~al.}{2009}]{ONeill2009}
O'Neill S.~M.,  {De Young} D.~S.,   Jones T.~W.,  2009, \mn@doi [Astrophys. J.]
  {10.1063/1.3293072}, 694, 1317

\bibitem[\protect\citeauthoryear{Ogorzalek et~al.,}{Ogorzalek
  et~al.}{2017}]{Ogorzalek2017a}
Ogorzalek A.,  et~al., 2017, \mn@doi [Mon. Not. R. Astron. Soc.]
  {10.1093/MNRAS/STX2030}, 472, 1659

\bibitem[\protect\citeauthoryear{Olivares et~al.,}{Olivares
  et~al.}{2019}]{Olivares2019}
Olivares V.,  et~al., 2019, \mn@doi [Astron. Astrophys.]
  {10.1051/0004-6361/201935350}, 631, A22

\bibitem[\protect\citeauthoryear{Ota \& Yoshida}{Ota \&
  Yoshida}{2016}]{Ota2016}
Ota N.,  Yoshida H.,  2016, \mn@doi [Publ. Astron. Soc. Japan]
  {10.1093/pasj/psv128}, 68, S19

\bibitem[\protect\citeauthoryear{Ota et~al.,}{Ota et~al.}{2007}]{Ota2007}
Ota N.,  et~al., 2007, \mn@doi [Prog. Theor. Phys. Suppl.]
  {10.1143/PTPS.169.25}, 59, S351

\bibitem[\protect\citeauthoryear{Pakmor, Springel, Bauer, Mocz, Munoz, Ohlmann,
  Schaal  \& Zhu}{Pakmor et~al.}{2016a}]{Pakmor2016}
Pakmor R.,  Springel V.,  Bauer A.,  Mocz P.,  Munoz D.~J.,  Ohlmann S.~T.,
  Schaal K.,   Zhu C.,  2016a, \mn@doi [Mon. Not. R. Astron. Soc.]
  {10.1093/mnras/stv2380}, 455, 1134

\bibitem[\protect\citeauthoryear{Pakmor, Pfrommer, Simpson, Kannan  \&
  Springel}{Pakmor et~al.}{2016b}]{Pakmor2016a}
Pakmor R.,  Pfrommer C.,  Simpson C.~M.,  Kannan R.,   Springel V.,  2016b,
  \mn@doi [Mon. Not. R. Astron. Soc.] {10.1093/mnras/stw1761}, 462, 2603

\bibitem[\protect\citeauthoryear{Pfrommer}{Pfrommer}{2013}]{Pfrommer2013}
Pfrommer C.,  2013, \mn@doi [Astrophys. J.] {10.1088/0004-637X/779/1/10}, 779,
  10

\bibitem[\protect\citeauthoryear{Pfrommer, Pakmor, Schaal, Simpson  \&
  Springel}{Pfrommer et~al.}{2017}]{Pfrommer2017}
Pfrommer C.,  Pakmor R.,  Schaal K.,  Simpson C.~M.,   Springel V.,  2017,
  \mn@doi [Mon. Not. R. Astron. Soc.] {10.1093/mnras/stw2941}, 465, 4500

\bibitem[\protect\citeauthoryear{Pinto et~al.,}{Pinto et~al.}{2015}]{Pinto2015}
Pinto C.,  et~al., 2015, \mn@doi [Astron. Astrophys.]
  {10.1051/0004-6361/201425278}, 575, A38

\bibitem[\protect\citeauthoryear{Prasad, Sharma  \& Babul}{Prasad
  et~al.}{2018}]{Prasad2018}
Prasad D.,  Sharma P.,   Babul A.,  2018, \mn@doi [Astrophys. J.]
  {10.3847/1538-4357/aacce8}, 863, 62

\bibitem[\protect\citeauthoryear{Rafferty, McNamara, Nulsen  \& Wise}{Rafferty
  et~al.}{2006}]{Rafferty2006}
Rafferty D.~A.,  McNamara B.~R.,  Nulsen P. E.~J.,   Wise M.~W.,  2006, \mn@doi
  [Astrophys. J.] {10.1086/507672}, 652, 216

\bibitem[\protect\citeauthoryear{Reiprich \& Bohringer}{Reiprich \&
  Bohringer}{2002}]{Reiprich2002}
Reiprich T.~H.,  Bohringer H.,  2002, \mn@doi [Astrophys. J.] {10.1086/338753},
  567, 716

\bibitem[\protect\citeauthoryear{Reynolds, Balbus  \& Schekochihin}{Reynolds
  et~al.}{2015}]{Reynolds2015}
Reynolds C.~S.,  Balbus S.~A.,   Schekochihin A.~A.,  2015, \mn@doi [Astrophys.
  J.] {10.1088/0004-637X/815/1/41}, 815, 41

\bibitem[\protect\citeauthoryear{Roh, Ryu, Kang, Ha  \& Jang}{Roh
  et~al.}{2019}]{Roh2019}
Roh S.,  Ryu D.,  Kang H.,  Ha S.,   Jang H.,  2019, \mn@doi [Astrophys. J.]
  {10.3847/1538-4357/ab3aff}, 883, 138

\bibitem[\protect\citeauthoryear{Rudnick \& Blundell}{Rudnick \&
  Blundell}{2003}]{Rudnick2003}
Rudnick L.,  Blundell K.~M.,  2003, \mn@doi [Astrophys. J.] {10.1086/373891},
  588, 143

\bibitem[\protect\citeauthoryear{Russell et~al.,}{Russell
  et~al.}{2019}]{Russell2019}
Russell H.~R.,  et~al., 2019, \mn@doi [Mon. Not. R. Astron. Soc.]
  {10.1093/mnras/stz2719}, 490, 3025

\bibitem[\protect\citeauthoryear{Ruszkowski, Yang  \& Reynolds}{Ruszkowski
  et~al.}{2017}]{Ruszkowski2017a}
Ruszkowski M.,  Yang H. Y.~K.,   Reynolds C.~S.,  2017, \mn@doi [Astrophys. J.]
  {10.3847/1538-4357/aa79f8}, 844, 13

\bibitem[\protect\citeauthoryear{Ryu, Kang, Cho  \& Das}{Ryu
  et~al.}{2008}]{Ryu2008}
Ryu D.,  Kang H.,  Cho J.,   Das S.,  2008, \mn@doi [Science (80-. ).]
  {10.1126/science.1154923}, 320, 909

\bibitem[\protect\citeauthoryear{Salom{\'{e}} et~al.,}{Salom{\'{e}}
  et~al.}{2006}]{Salome2006}
Salom{\'{e}} P.,  et~al., 2006, \mn@doi [Astron. Astrophys.]
  {10.1051/0004-6361:20054745}, 454, 437

\bibitem[\protect\citeauthoryear{Sanders \& Fabian}{Sanders \&
  Fabian}{2013}]{Sanders2013}
Sanders J.~S.,  Fabian A.~C.,  2013, \mn@doi [Mon. Not. R. Astron. Soc.]
  {10.1093/mnras/sts543}, 429, 2727

\bibitem[\protect\citeauthoryear{Sanders, Fabian, Smith  \& Peterson}{Sanders
  et~al.}{2010}]{Sanders2010}
Sanders J.~S.,  Fabian A.~C.,  Smith R.~K.,   Peterson J.~R.,  2010, \mn@doi
  [Mon. Not. R. Astron. Soc. Lett.] {10.1111/j.1745-3933.2009.00789.x}, 402,
  L11

\bibitem[\protect\citeauthoryear{Schekochihin \& Cowley}{Schekochihin \&
  Cowley}{2006}]{Schekochihin2006}
Schekochihin A.~A.,  Cowley S.~C.,  2006, \mn@doi [Phys. Plasmas]
  {10.1063/1.2179053}, 13

\bibitem[\protect\citeauthoryear{Schober, Schleicher, Federrath, Bovino  \&
  Klessen}{Schober et~al.}{2015}]{Schober2015}
Schober J.,  Schleicher D.~R.,  Federrath C.,  Bovino S.,   Klessen R.~S.,
  2015, \mn@doi [Phys. Rev. E - Stat. Nonlinear, Soft Matter Phys.]
  {10.1103/PhysRevE.92.023010}, 92, 023010

\bibitem[\protect\citeauthoryear{Schuecker, Finoguenov, Miniati, Boehringer  \&
  Briel}{Schuecker et~al.}{2004}]{Schuecker2004}
Schuecker P.,  Finoguenov A.,  Miniati F.,  Boehringer H.,   Briel U.~G.,
  2004, Astron. Astrophys., 426, 387

\bibitem[\protect\citeauthoryear{Sharma, Chandran, Quataert  \& Parrish}{Sharma
  et~al.}{2009}]{Sharma2009}
Sharma P.,  Chandran B. D.~G.,  Quataert E.,   Parrish I.~J.,  2009, \mn@doi
  [ApJ] {10.1088/0004-637X/699/1/348}, 699, 348

\bibitem[\protect\citeauthoryear{Simionescu et~al.,}{Simionescu
  et~al.}{2019}]{Simionescu2019}
Simionescu A.,  et~al., 2019, \mn@doi [Space Sci. Rev.]
  {10.1007/s11214-019-0590-1}, 215

\bibitem[\protect\citeauthoryear{Smith, Brickhouse, Liedahl  \& Raymond}{Smith
  et~al.}{2001}]{Smith2001}
Smith R.~K.,  Brickhouse N.~S.,  Liedahl D.~A.,   Raymond J.~C.,  2001, \mn@doi
  [Astrophys. J.] {10.1086/322992}, 556, L91

\bibitem[\protect\citeauthoryear{Sparre, Pfrommer  \& Ehlert}{Sparre
  et~al.}{2020}]{Sparre2020}
Sparre M.,  Pfrommer C.,   Ehlert K.,  2020, \mn@doi [Mon. Not. R. Astron.
  Soc.] {10.1093/mnras/staa3177}, 4281, 4261

\bibitem[\protect\citeauthoryear{Springel}{Springel}{2010}]{Springel2010}
Springel V.,  2010, \mn@doi [Mon. Not. R. Astron. Soc.]
  {10.1111/j.1365-2966.2009.15715.x}, 401, 791

\bibitem[\protect\citeauthoryear{Subramanian}{Subramanian}{1999}]{Subramanian1999}
Subramanian K.,  1999, Phys. Rev. Lett., 83, 15

\bibitem[\protect\citeauthoryear{Sugawara, Takizawa  \& Nakazawa}{Sugawara
  et~al.}{2009}]{Sugawara2009}
Sugawara C.,  Takizawa M.,   Nakazawa K.,  2009, \mn@doi [Publ. Astron. Soc.
  Japan] {10.1093/pasj/61.6.1293}, 61, 1293

\bibitem[\protect\citeauthoryear{Tamura, Hayashida, Ueda  \& Nagai}{Tamura
  et~al.}{2011}]{Tamura2011}
Tamura T.,  Hayashida K.,  Ueda S.,   Nagai M.,  2011, \mn@doi [Publ. Astron.
  Soc. Japan] {10.1093/pasj/63.sp3.s1009}, 63, S1009

\bibitem[\protect\citeauthoryear{Tamura et~al.,}{Tamura
  et~al.}{2014}]{Tamura2014}
Tamura T.,  et~al., 2014, \mn@doi [Astrophys. J.] {10.1088/0004-637X/782/1/38},
  782, 38

\bibitem[\protect\citeauthoryear{Tang \& Churazov}{Tang \&
  Churazov}{2017}]{Tang2017}
Tang X.,  Churazov E.,  2017, \mn@doi [Mon. Not. R. Astron. Soc.]
  {10.1093/mnras/stx590}, 468, 3516

\bibitem[\protect\citeauthoryear{Tchekhovskoy \& Bromberg}{Tchekhovskoy \&
  Bromberg}{2016}]{Tchekhovskoy2016}
Tchekhovskoy A.,  Bromberg O.,  2016, \mn@doi [Mon. Not. R. Astron. Soc.]
  {10.1093/mnrasl/slw064}, 461, L46

\bibitem[\protect\citeauthoryear{Thomas \& Pfrommer}{Thomas \&
  Pfrommer}{2019}]{Thomas2019}
Thomas T.,  Pfrommer C.,  2019, \mn@doi [Mon. Not. R. Astron. Soc.]
  {10.1093/mnras/stz263}, 485, 2977

\bibitem[\protect\citeauthoryear{Thomas, Pfrommer  \& En{\ss}lin}{Thomas
  et~al.}{2020}]{Thomas2020}
Thomas T.,  Pfrommer C.,   En{\ss}lin T.,  2020, \mn@doi [Astrophys. J. Lett.]
  {10.3847/2041-8213/ab7237}, 890, L18

\bibitem[\protect\citeauthoryear{Tremblay et~al.,}{Tremblay
  et~al.}{2018}]{Tremblay2018}
Tremblay G.~R.,  et~al., 2018, \mn@doi [Astrophys. J.]
  {10.3847/1538-4357/aad6dd}, 865, 13

\bibitem[\protect\citeauthoryear{Vacca, Murgia, Govoni, Feretti, Giovannini,
  Perley  \& Taylor}{Vacca et~al.}{2012}]{Vacca2012}
Vacca V.,  Murgia M.,  Govoni F.,  Feretti L.,  Giovannini G.,  Perley R.~a.,
  Taylor G.~B.,  2012, \mn@doi [Astron. Astrophys.]
  {10.1051/0004-6361/201116622}, 540, A38

\bibitem[\protect\citeauthoryear{Vacca, Murgia, Govoni, En{\ss}lin, Oppermann,
  Feretti, Giovannini  \& Loi}{Vacca et~al.}{2018}]{Vacca2018}
Vacca V.,  Murgia M.,  Govoni F.,  En{\ss}lin T.,  Oppermann N.,  Feretti L.,
  Giovannini G.,   Loi F.,  2018, \mn@doi [Galaxies] {10.3390/galaxies6040142},
  6, 1

\bibitem[\protect\citeauthoryear{Vantyghem et~al.,}{Vantyghem
  et~al.}{2019}]{Vantyghem2019}
Vantyghem A.~N.,  et~al., 2019, \mn@doi [Astrophys. J.]
  {10.3847/1538-4357/aaf1b4}, 870, 57

\bibitem[\protect\citeauthoryear{Vazza, Brunetti, Br{\"{u}}ggen  \&
  Bonafede}{Vazza et~al.}{2018}]{Vazza2018a}
Vazza F.,  Brunetti G.,  Br{\"{u}}ggen M.,   Bonafede A.,  2018, \mn@doi [Mon.
  Not. R. Astron. Soc.] {10.1093/mnras/stx2830}, 474, 1672

\bibitem[\protect\citeauthoryear{Vogt \& En{\ss}lin}{Vogt \&
  En{\ss}lin}{2005}]{Vogt2005}
Vogt C.,  En{\ss}lin T.~A.,  2005, \mn@doi [Astron. Astrophys.]
  {10.1051/0004-6361:20041839}, 434, 67

\bibitem[\protect\citeauthoryear{Voit, Meece, Li, Shea, Bryan  \& Donahue}{Voit
  et~al.}{2017}]{Voit2017}
Voit G.~M.,  Meece G.,  Li Y.,  Shea B. W.~O.,  Bryan G.~L.,   Donahue M.,
  2017, \mn@doi [Astrophys. J.] {10.3847/1538-4357/aa7d04}, 845, 80

\bibitem[\protect\citeauthoryear{Walker, Sanders  \& Fabian}{Walker
  et~al.}{2015}]{Walker2015}
Walker S.~A.,  Sanders J.~S.,   Fabian A.~C.,  2015, \mn@doi [Mon. Not. R.
  Astron. Soc.] {10.1093/mnras/stv1929}, 453, 3699

\bibitem[\protect\citeauthoryear{Walker, Sanders  \& Fabian}{Walker
  et~al.}{2018}]{Walker2018a}
Walker S.~A.,  Sanders J.~S.,   Fabian A.~C.,  2018, \mn@doi [Mon. Not. R.
  Astron. Soc.] {10.1093/MNRAS/STY2390}, 481, 1718

\bibitem[\protect\citeauthoryear{Weinberger, Ehlert, Pfrommer, Pakmor  \&
  Springel}{Weinberger et~al.}{2017}]{Weinberger2017}
Weinberger R.,  Ehlert K.,  Pfrommer C.,  Pakmor R.,   Springel V.,  2017,
  \mn@doi [Mon. Not. R. Astron. Soc.] {10.1093/mnras/stx1409}, 470, 4530

\bibitem[\protect\citeauthoryear{Werner, Zhuravleva, Churazov, Simionescu,
  Allen, Forman, Jones  \& Kaastra}{Werner et~al.}{2009}]{Werner2009}
Werner N.,  Zhuravleva I.,  Churazov E.,  Simionescu A.,  Allen S.~W.,  Forman
  W.,  Jones C.,   Kaastra J.~S.,  2009, \mn@doi [Mon. Not. R. Astron. Soc.]
  {10.1111/j.1365-2966.2009.14860.x}, 398, 23

\bibitem[\protect\citeauthoryear{Werner et~al.,}{Werner
  et~al.}{2014}]{Werner2014a}
Werner N.,  et~al., 2014, \mn@doi [Mon. Not. R. Astron. Soc.]
  {10.1093/mnras/stu006}, 439, 2291

\bibitem[\protect\citeauthoryear{Wiener, Pfrommer  \& {Peng Oh}}{Wiener
  et~al.}{2017}]{Wiener2017}
Wiener J.,  Pfrommer C.,   {Peng Oh} S.,  2017, \mn@doi [Mon. Not. R. Astron.
  Soc.] {10.1093/mnras/stx127}, 467, 906

\bibitem[\protect\citeauthoryear{Wilms, Allen  \& McCray}{Wilms
  et~al.}{2000}]{Wilms2000}
Wilms J.,  Allen A.,   McCray R.,  2000, \mn@doi [Astrophys. J.]
  {10.1086/317016}, 542, 914

\bibitem[\protect\citeauthoryear{Xu et~al.,}{Xu et~al.}{2002}]{Xu2002}
Xu H.,  et~al., 2002, \mn@doi [Astrophys. J.] {10.1086/342828}, 579, 600

\bibitem[\protect\citeauthoryear{Xu, Li, Collins, Li  \& Norman}{Xu
  et~al.}{2009}]{Xu2009}
Xu H.,  Li H.,  Collins D.~C.,  Li S.,   Norman M.~L.,  2009, \mn@doi
  [Astrophys. J.] {10.1088/0004-637X/698/1/L14}, 698, L14

\bibitem[\protect\citeauthoryear{Yang \& Reynolds}{Yang \&
  Reynolds}{2016}]{Yang2016}
Yang H.-Y.~K.,  Reynolds C.~S.,  2016, \mn@doi [Astrophys. J.]
  {10.3847/0004-637x/829/2/90}, 829, 90

\bibitem[\protect\citeauthoryear{Yuan, Gan, Narayan, Sadowski, Bu  \& Bai}{Yuan
  et~al.}{2015}]{Yuan2015}
Yuan F.,  Gan Z.,  Narayan R.,  Sadowski A.,  Bu D.,   Bai X.-n.,  2015,
  \mn@doi [Astrophys. J.] {10.1088/0004-637X/804/2/101}, 804, 101

\bibitem[\protect\citeauthoryear{Zhuravleva et~al.,}{Zhuravleva
  et~al.}{2014}]{Zhuravleva2014a}
Zhuravleva I.,  et~al., 2014, \mn@doi [Nature] {10.1038/nature13830}, 515, 85

\bibitem[\protect\citeauthoryear{Zhuravleva et~al.,}{Zhuravleva
  et~al.}{2016}]{Zhuravleva2016a}
Zhuravleva I.,  et~al., 2016, \mn@doi [Mon. Not. R. Astron. Soc.]
  {10.1093/mnras/stw520}, 458, 2902

\bibitem[\protect\citeauthoryear{Zhuravleva, Allen, Mantz  \&
  Werner}{Zhuravleva et~al.}{2018}]{Zhuravleva2018}
Zhuravleva I.,  Allen S.~W.,  Mantz A.,   Werner N.,  2018, \mn@doi [Astrophys.
  J.] {10.3847/1538-4357/aadae3}, 865, 53

\bibitem[\protect\citeauthoryear{ZuHone \& Hallman}{ZuHone \&
  Hallman}{2016}]{ZuHone2016b}
ZuHone J.~A.,  Hallman E.~J.,  2016, Astrophys. Source Code Libr.

\bibitem[\protect\citeauthoryear{ZuHone, Miller, Bulbul  \& Zhuravleva}{ZuHone
  et~al.}{2018}]{ZuHone2018}
ZuHone J.,  Miller E.~D.,  Bulbul E.,   Zhuravleva I.,  2018, \mn@doi
  [Astrophys. J.] {10.3847/1538-4357/aaa4b3}, 853, 180

\bibitem[\protect\citeauthoryear{Zweibel}{Zweibel}{2013}]{Zweibel2013}
Zweibel E.~G.,  2013, \mn@doi [Phys. Plasmas] {10.1063/1.4807033}, 20, 055501

\bibitem[\protect\citeauthoryear{de Gasperin et~al.,}{de~Gasperin
  et~al.}{2012}]{DeGasperin2012}
de Gasperin F.,  et~al., 2012, \mn@doi [Astron. Astrophys.]
  {10.1051/0004-6361/201220209}, 547, A56

\bibitem[\protect\citeauthoryear{van Weeren, de Gasperin, Akamatsu,
  Br{\"{u}}ggen, Feretti, Kang, Stroe  \& Zandanel}{van Weeren
  et~al.}{2019}]{VanWeeren2019}
van Weeren R.~J.,  de Gasperin F.,  Akamatsu H.,  Br{\"{u}}ggen M.,  Feretti
  L.,  Kang H.,  Stroe A.,   Zandanel F.,  2019, \mn@doi [Space Sci. Rev.]
  {10.1007/s11214-019-0584-z}, 215

\makeatother
\end{thebibliography}

%%%%%%%%%%%%%%%%%%%%%%%%%%%%%%%%%%%%%%%%%%%%%%%%%%

%%%%%%%%%%%%%%%%% APPENDICES %%%%%%%%%%%%%%%%%%%%%

\appendix

\section{Varying magnetic field parameters}
\label{sec:varymagneticfieldstrength}
\begin{figure*}
\centering
\includegraphics[trim=0.2cm .2cm 0.2cm .2cm,clip=true, width=0.8\textwidth]{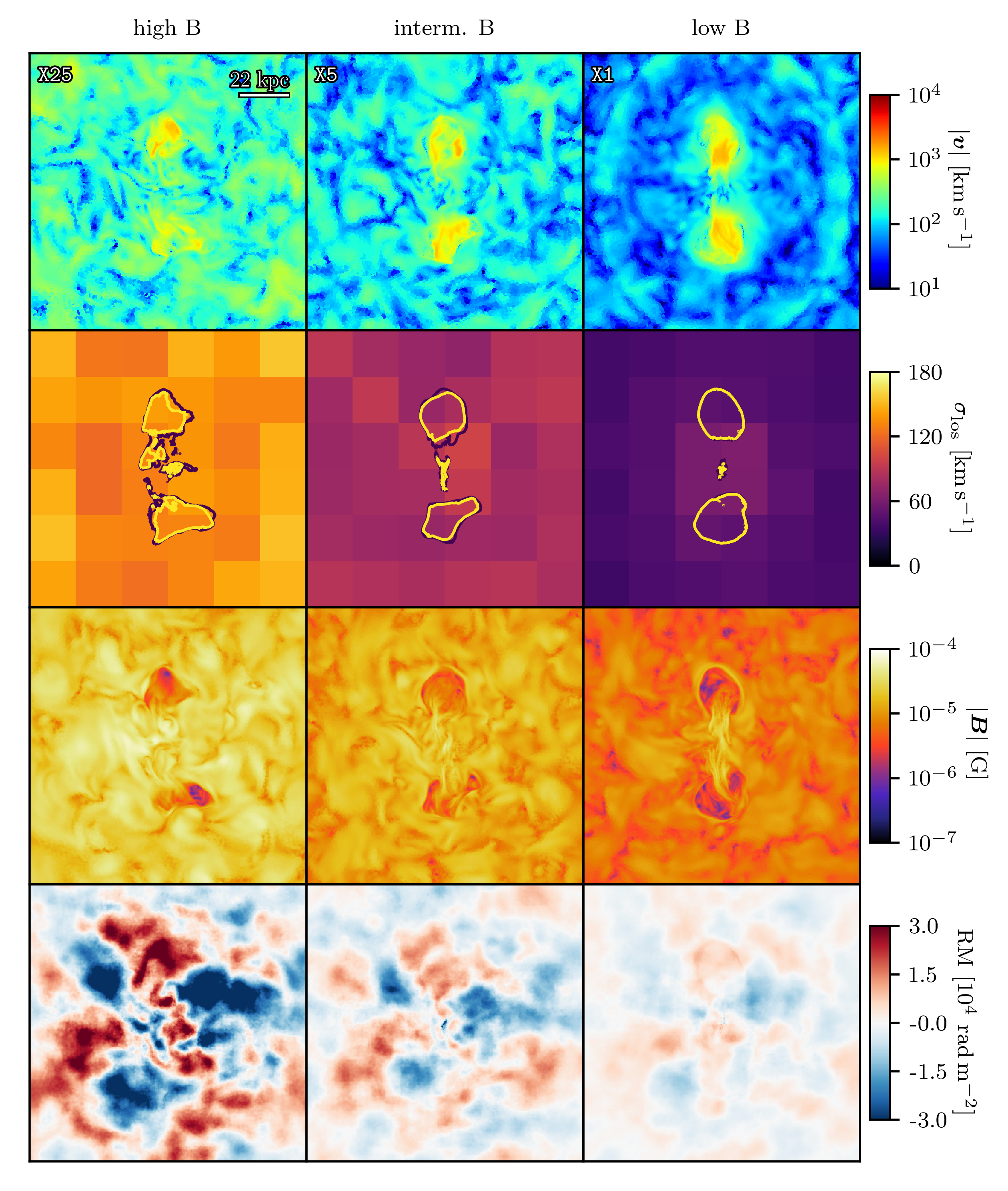}
\caption{From left to right, we compare runs \texttt{X25}, \texttt{X5}, \texttt{X1} with varying magnetic field strengths $X_B=0.25,\,0.05,\,0.01$, respectively. A higher magnetic fields strength provides a stronger tension force that induces higher velocities and velocity dispersion in the ICM. Both \texttt{X25} and \texttt{X5} induce a velocity dispersion consistent with \textit{Hitomi} measurements, \texttt{X25} is at the upper end and \texttt{X5} thereby preferred. See Figure \ref{fig:hitomi_compare} for details on shown quantities.}
    \label{fig:hitomi_magneticvaried}
\end{figure*}

We varied the initial magnetic field strength in our simulations and depict the resulting velocity dispersion in Figure \ref{fig:hitomi_magneticvaried}. The magnetic-to-thermal pressure ratio in the ICM $X_{B,\mathrm{ICM}}$ decreases as $X_{B,\mathrm{ICM}}=0.25,\,0.05,\,0.01$ from left to right. Consequently, the velocity dispersion decreases as $\gtrsim150\,\mathrm{km}\,\mathrm{s}^{-1}$,  $\sim100\,\mathrm{km}\,\mathrm{s}^{-1}$ and $\lesssim 60\,\mathrm{km}\,\mathrm{s}^{-1}$, respectively. The lowest magnetic field run \texttt{X1} cannot stir the ICM sufficiently, to reach the velocity dispersion observed by \textit{Hitomi}. Both \texttt{X25} and \texttt{X5} produce a velocity dispersion that is consistent with \textit{Hitomi} measurements. However, we emphasize that cosmological simulations are necessary to follow the evolution of the magnetic dynamo and obtain self-consistent velocity fields. 

\begin{figure*}
\centering
\includegraphics[trim=0.2cm .2cm 0.2cm .2cm,clip=true, width=0.76\textwidth]{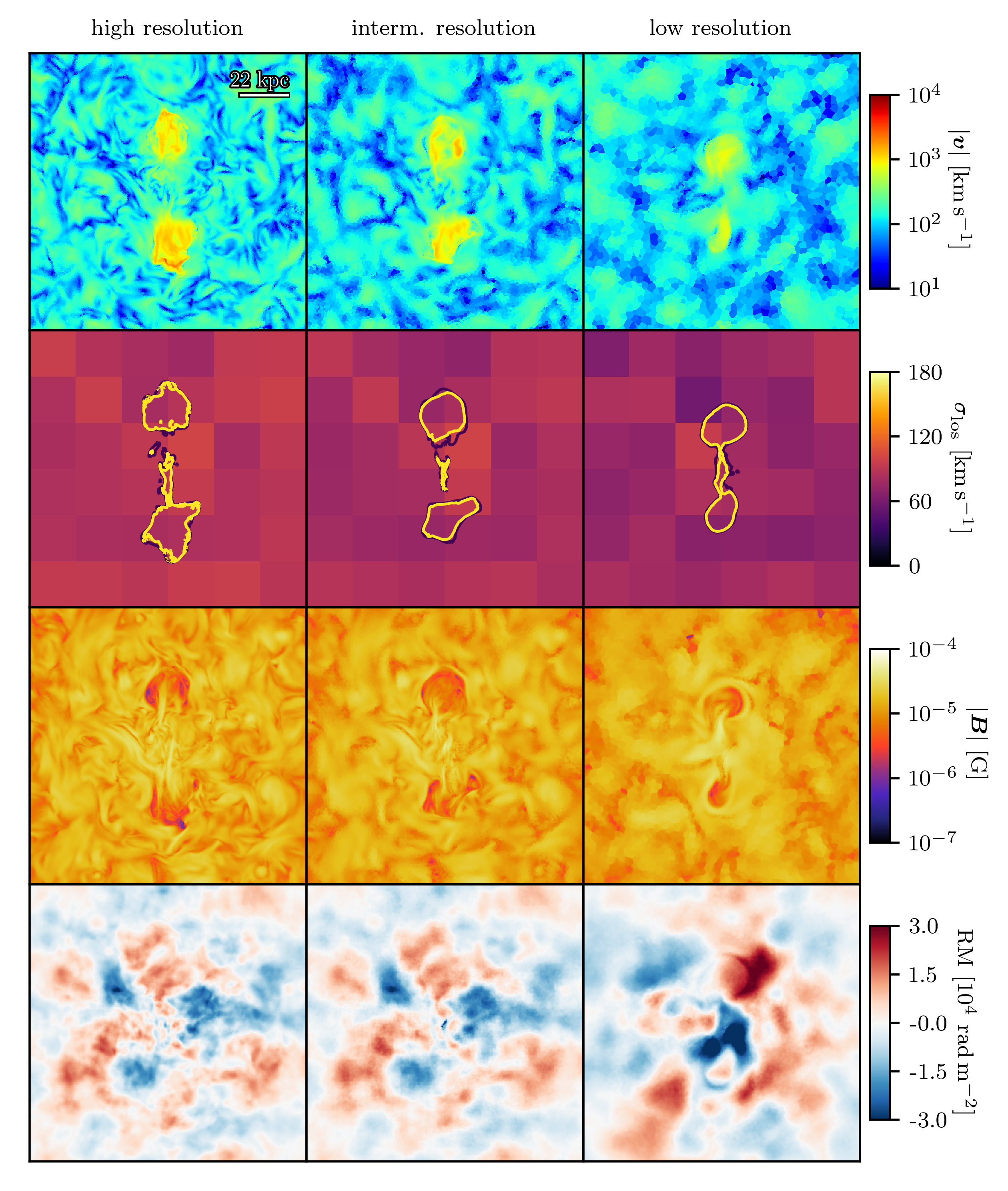}
\caption{From left to right, we compare simulations at high, intermediate and low resolution, respectively. The loss of small scale structures leads to an increase in RM for the low resolution run as depolarization is reduced. However, overall features in runs at high and intermediate resolution show convergence. See Figure \ref{fig:hitomi_compare} for details on shown quantities.}
    \label{fig:hitomi_resolutionvaried}
\end{figure*}

In addition, we show simulations at decreasing resolution from left to right in Figure \ref{fig:hitomi_resolutionvaried}. The velocity dispersion decreases slightly with resolution. This is likely due to the increased numerical diffusivity in the lower resolution runs, which decrease the effectiveness of the stirring on longer timescales. The RMs show higher maxima in the low resolution run. Intermittent magnetic field strengths are less resolved so that the cancelling of RM is reduced.

\begin{figure}
\centering
\includegraphics[trim=3.cm 0.5cm 3.cm 1.5cm,clip=true, width=0.47\textwidth]{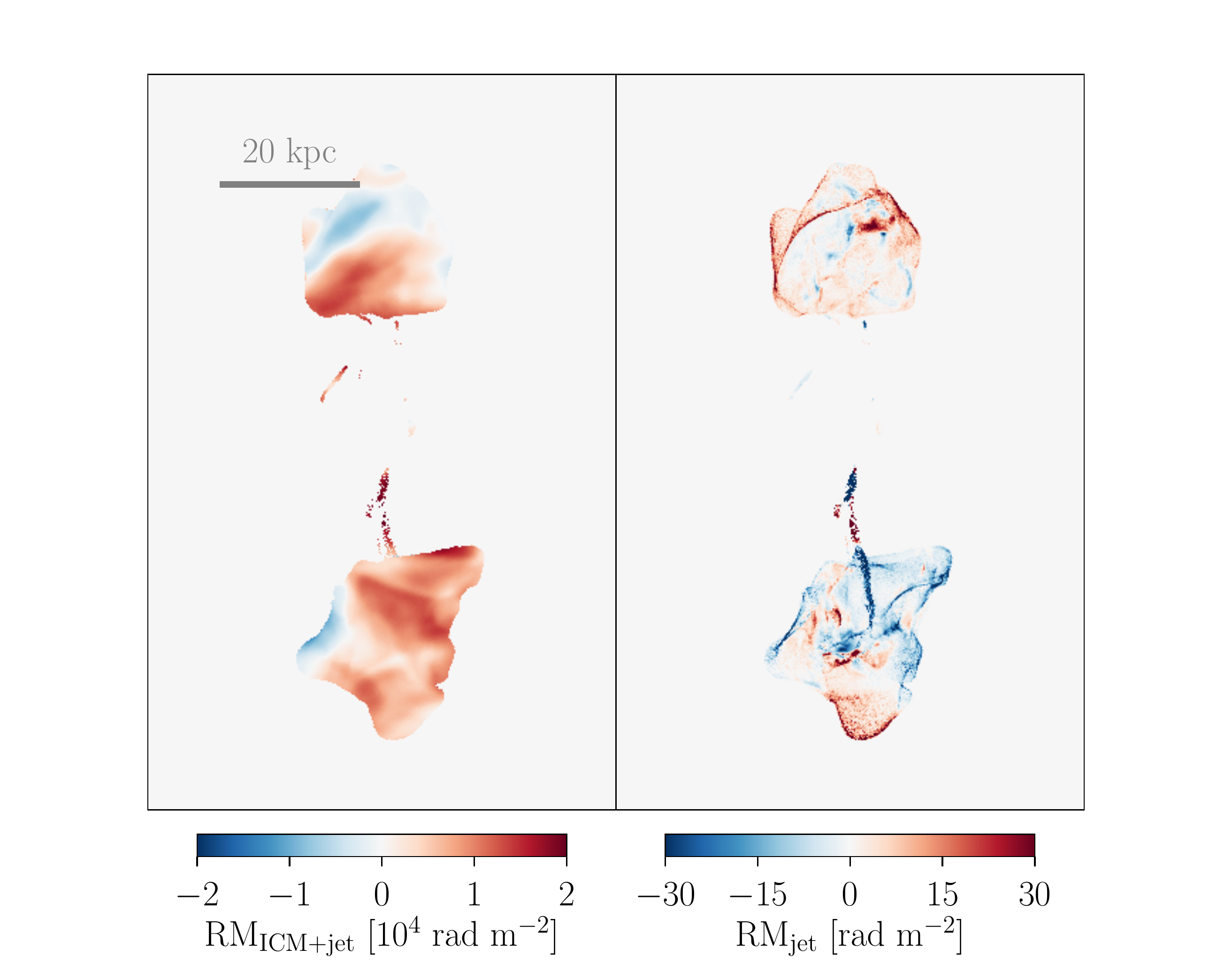}
\caption{We show the RM of the fiducial run at $50\ \mathrm{Myr}$ for varying sources of emission. On the left, we include both the ICM and AGN as the source. On the right, we show the contribution from jet only ($X_\mathrm{jet}>10^{-3}$). The ICM contribution dominates the RM signal by at least two orders of magnitude. The enhancement in parts of the bubbles' rim and some filaments is an artifact of the tagging process of the lobes.}
    \label{fig:RM_compare}
\end{figure}

Turning our attention now to the influence of jet magnetic fields on the overall RM, we see that RM is dominated by the contributions from the ICM. In Figure \ref{fig:RM_compare} we compare the total RM from ICM and jet (left) with the RM from the jet only (right). Some rims of the bubble and sparse filaments show relatively high signal. Here, only a few cells exceed the threshold in $X_\mathrm{jet}$ and thereby suffer from minimal depolarization. Comparing magnitudes, we see that the lobes contribute at least two orders of magnitude less signal compared to the ICM. This is consistent with observations, that would otherwise generally suffer from considerable beam polarization \citep[e.g.,][]{Han2017}. Note however, that some sources show evidence for a dominating contribution from locally compressed ICM close to the lobes \citep{Rudnick2003,Guidetti2012}. 

\section{Jet resolution study}
\label{sec:jetdistance}
\begin{figure}
\centering
\includegraphics[trim=.1cm .1cm 0.cm .cm,clip=true, width=0.5\textwidth]{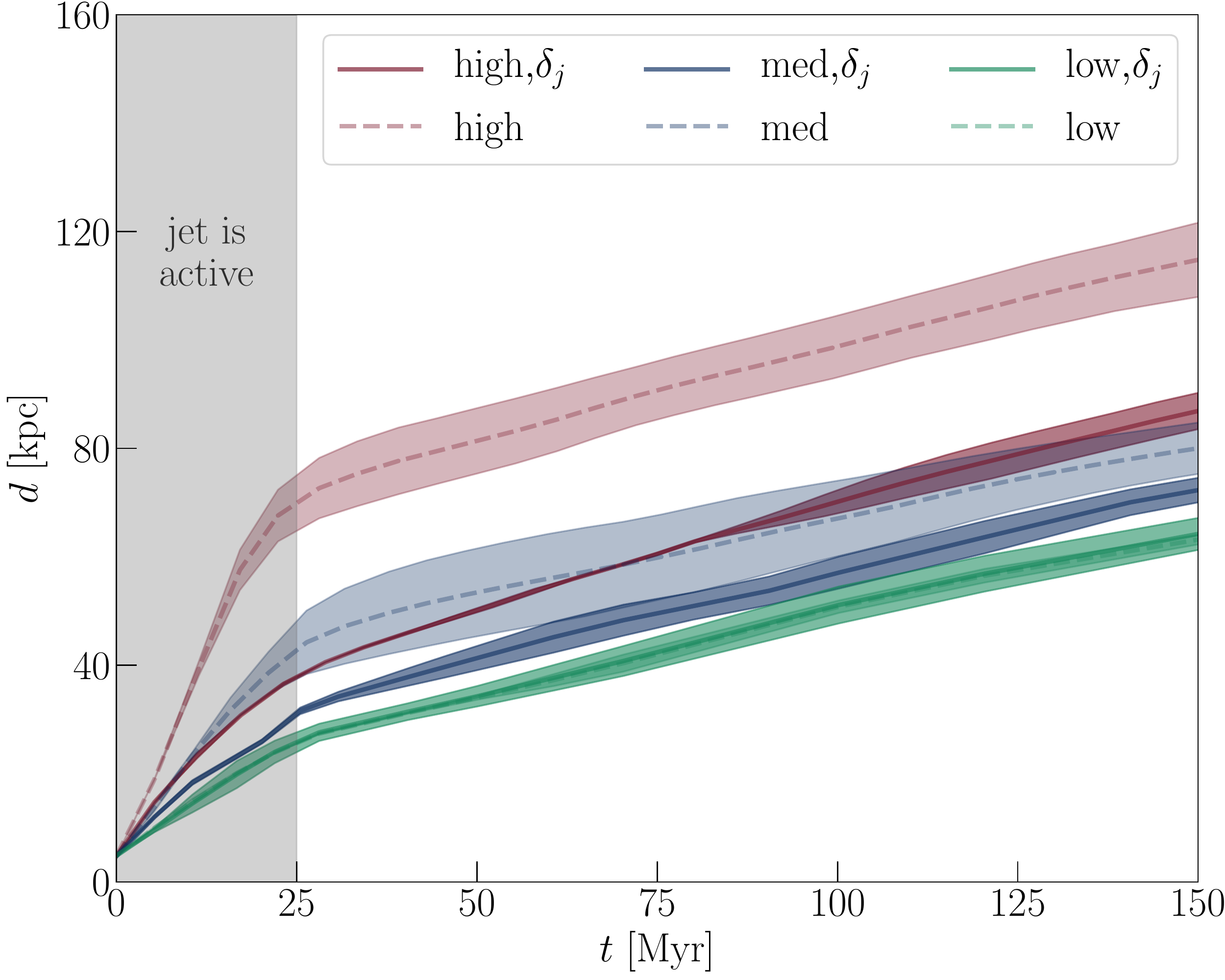}
\caption{Average distance of the jets at low, intermediate and high resolution (shown with different colours) as a function of time. We compare our new model that includes an opening angle of $\delta_j=30^\circ$ (solid lines) to simulations with our previous model ($\delta_j=0^\circ$, dashed lines). The errorbars indicate the distance of the individual jets. Including an opening angle leads to improved numerical convergence of the jet distance travelled, especially for jets at high resolution.}
    \label{fig:jetangle_distance_compare}
\end{figure}

To improve the numerical convergence with respect to the distance traveled of our jets, we introduce an opening angle $\delta_j$ to the model. As discussed in Section~\ref{sec:jetmodel}, half of the momentum is injected with an angle smaller than $10$~degree from the jet axis, while the maximum angle is $30$~degree. While the jet initially fans out, it is almost immediately collimated by the pressure of the ambient ICM. This leads to an overall broadening of the jet where better resolved jets are affected more. Thereby better convergence of jet distance is obtained. In Figure \ref{fig:jetangle_distance_compare}, we contrast the bubble distance with and without opening angle as a function of time for simulations at three different resolutions. Simulations with an opening angle of $\delta_j=30^\circ$ converge more to similar distances than those without opening angle ($\delta_j=0^\circ$).

\section{Synthetic X-ray observations}
\label{app:xray}

\begin{figure*}
\centering
\includegraphics[trim=2cm 1cm 0.2cm 0.8cm, clip=true, width = 0.7\textwidth]{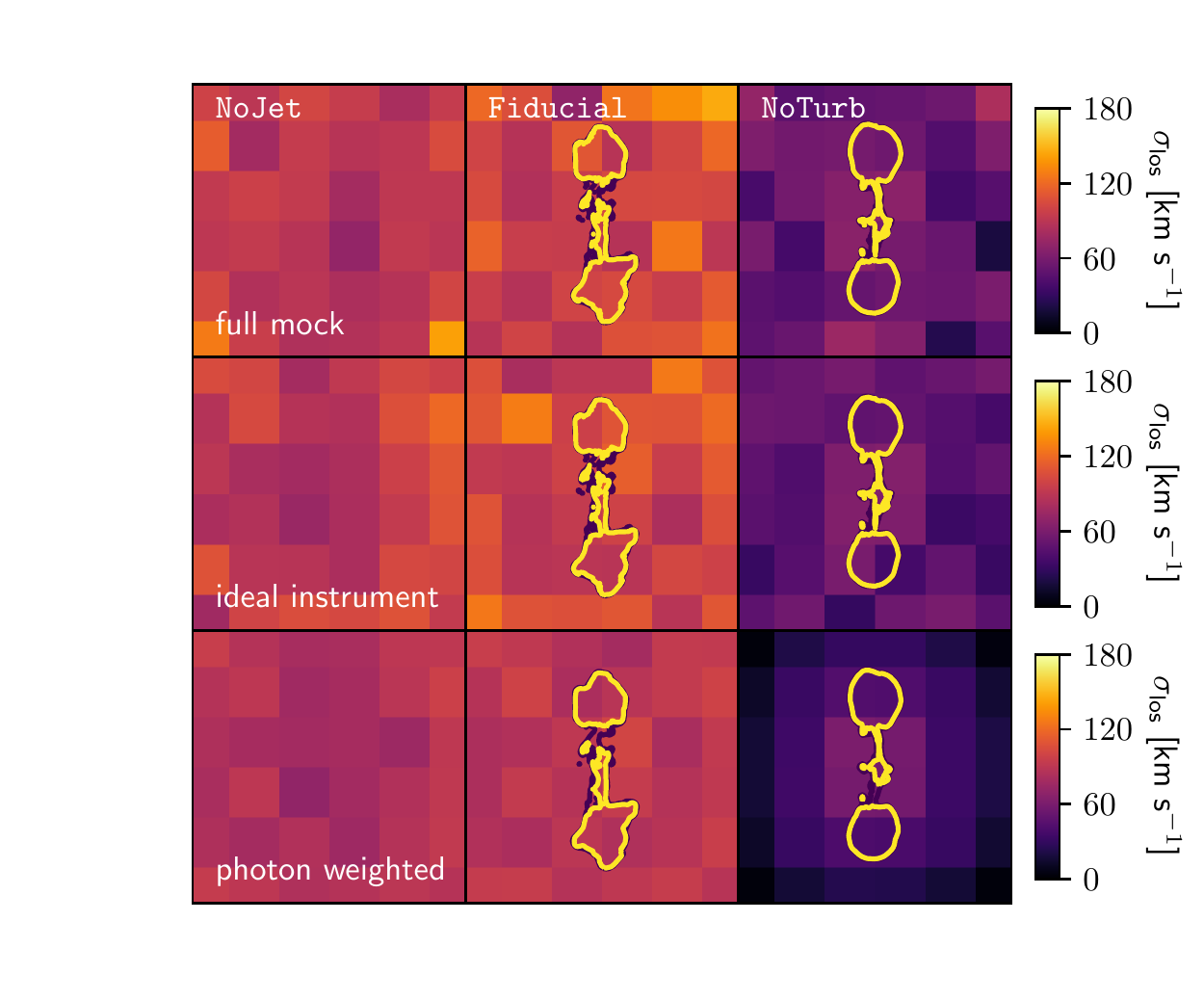}
\caption{Velocity dispersion maps from mock Xrism observations (top), fits to the emitted photon spectrum (center) and 
the emission weighted velocity dispersion (bottom).}
\label{fig:mock_compare}
\end{figure*}

Figure~\ref{fig:mock_compare} shows velocity dispersion maps using different methods from a full synthetic observation (top row), a fit to photon spectrum, i.e. synthetic observation without instrument response (middle row), and a $2-12$~keV emission weighted velocity dispersion. While there is a systematic increase in velocity dispersion originating from a higher temperatures at the outskirts of the projection, the instrumental effect mainly introduces scatter on a pixel by pixel basis.

\bsp
\label{lastpage}
\end{document}